\documentclass[printer]{aa}  
\usepackage[english]{babel}
\usepackage[T1]{fontenc}
\usepackage{graphicx}
\usepackage{txfonts}
\usepackage{hyperref}

\usepackage{newtxtext,newtxmath}
\usepackage{longtable}
\usepackage{setspace}
\usepackage{booktabs}
\usepackage{wallpaper}
\usepackage{epstopdf}
\usepackage{eso-pic}
\usepackage{eqparbox}

\usepackage[singlelinecheck=off]{caption}
\usepackage{longtable}

\usepackage{ae,aecompl}

\usepackage{amsmath}	
\usepackage{amssymb}	

\def\cratio{$^{12}$C/$^{13}$C}
\def\nratio{$^{14}$N/$^{15}$N}

\newcommand{\Arrow}{\mathop{\longrightarrow}\limits}
\newcommand{\asec}{$^{\prime\prime}$}
\newcommand{\GG}[1]{}

\begin{document}

   \title{Carbon isotopic fractionation in molecular clouds}


   \author{L. Colzi\inst{1,2}
          \and
          O. Sipilä\inst{3}
          \and
          E. Roueff\inst{4}
          \and
          P. Caselli\inst{3}
          \and
          F. Fontani\inst{2}
          }

   \institute{Università degli studi di Firenze, Dipartimento di fisica e Astronomia, Via Sansone 1, 50019 Sesto Fiorentino, Italy\\
              \email{lcolzi.astro@gmail.com}
         \and
            INAF-Osservatorio Astrofisico di Arcetri, Largo E. Fermi 5, I-50125, Florence, Italy 
           \and 
           Max-Planck-Instit\"{u}t f\"{u}r extraterrestrische Physik, Giessenbachstrasse 1, D-85748, Garching bei M\"{u}nchen, Germany
           \and
          Sorbonne Université, Observatoire de Paris, Université PSL, CNRS, LERMA, 92190 Meudon, France   
           }

\date{Received 24 April 2020 / Accepted 4 June 2020}

 
  \abstract
   {C-fractionation has been studied from a theoretical point of view with different models of time-dependent chemistry, including both isotope-selective photodissociation and low-temperature isotopic exchange reactions.}
   {Recent chemical models predict that the latter may lead to a depletion of $^{13}$C in nitrile-bearing species, with \cratio\;ratios two times higher than the elemental abundance ratio of 68 in the local interstellar medium. Since the carbon isotopic ratio is commonly used to evaluate the \nratio\;ratios with the double-isotope method, it is important to study carbon fractionation in detail to avoid incorrect assumptions.}
   {In this work we implemented a gas-grain chemical model with new isotopic exchange reactions and investigated their introduction in the context of dense and cold molecular gas. In particular, we investigated the \cratio\;ratios of HNC, HCN, and CN using a grid of models, with temperatures and densities ranging from 10 to 50 K and 2$\times$10$^{3}$ to 2$\times$10$^{7}$ cm$^{-3}$, respectively.}
   {We suggest a possible $^{13}$C exchange through the $^{13}$C + C$_{3}$ $\rightarrow$ $^{12}$C +$^{13}$CC$_{2}$ reaction, which does not result in dilution, but rather in $^{13}$C enhancement, for molecules that are formed starting from atomic carbon. This effect is efficient in a range of time between the formation of CO and its freeze-out on grains. Furthermore, the parameter-space exploration shows, on average, that the \cratio\;ratios of nitriles are predicted to be a factor 0.8--1.9 different from the local \cratio\;of 68 for high-mass star-forming regions. This result also affects the \nratio\;ratio: a value of 330 obtained with the double-isotope method is predicted to vary in the range 260--630, up to 1150, depending on the physical conditions. Finally, we studied the \cratio\;ratios of nitriles by varying the cosmic-ray ionization rate, $\zeta$: the \cratio\;ratios increase with $\zeta$ because of secondary photons and cosmic-ray reactions.}
   {}
   

   \maketitle
   
%

\section{Introduction}
 \label{intro}
 
Isotopic fractionation is an important chemical process that occurs in interstellar clouds. It is the set of processes that distributes the less abundant stable isotopes of an element into other molecular species. Understanding isotopic abundances over a large range of scales, from terrestrial oceans, meteorites, planetary, and cometary atmospheres up to Galactic and extragalactic is crucial.
In particular the variation in isotopic ratios may give important information about the link between Solar system objects and Galactic interstellar environments (e.g. \citealt{caselliceccarelli2012}; \citealt{hily-blant2013a}; \citealt{hily-blant2013b}; \citealt{ceccarelli2014}; \citealt{fontani2015b}; \citealt{colzi18a}; \citealt{colzi18b}, \citealt{colzi2019}, and \citealt{fontani2020}).

One element for which fractionation is important is carbon. C-fractionation in dense interstellar clouds has been studied with different models of time-dependent chemistry. \citet{langer1984} introduced different isotopic exchange reactions in their model, as proposed by \citet{watson1976}. In particular, they proposed that C-fractionation in interstellar species is a result of the isotopic exchange reaction
\begin{equation}
\label{C-reaction1}
^{13}{\rm C}^{+} +\hspace{0.1cm}^{12}{\rm CO} \leftrightarrow \hspace{0.1cm}^{13}{\rm CO} + \hspace{0.1cm}^{12}{\rm C}^{+} + 35 \hspace{0.1cm}{\rm K}.
\end{equation}
They concluded that the lower the temperature, the higher the chemical fractionation of C-bearing species. In fact, this reaction leads to a $^{13}$C enhancement in CO and to a dilution in species formed from C$^{+}$ when all of the carbon is not fully locked in CO. \citet{smith1980}, under the suggestion of \citet{langer1978}, measured the rate coefficient for another isotopic exchange reaction:
\begin{equation}
\label{C-reaction2}
{\rm HCO}^{+} + \hspace{0.1cm}^{13}{\rm CO} \leftrightarrow {\rm H}^{13}{\rm CO}^{+} + {\rm CO} + \Delta E\hspace{0.1cm}{\rm K},
\end{equation}
with $\Delta E$=12$\pm$4. However, more recently, this reaction has been revised theoretically by \citet{mladenovic2014} who derived a value of 17.4 K from detailed zero-point energies (ZPE) calculations. This value is very similar to the one estimated by \citet{hennig1977} of 17$\pm$1 K. All of these reactions are very important in a low temperature ($\sim$20 K) environment.

In more recent works, fractionation effects due to the photodissociation of CO by ultraviolet (UV) photons, based on the theoretical studies by \citet{vandishoeck1988}, has been taken into account in chemical models. In particular, the selective photodissociation of $^{12}$CO is expected to be dominant in low density (< 10$^{2}$ cm$^{-3}$) environments, or denser regions with a strong radiation field. 
\citet{rollig2013} found, with their photodissociation region (PDR) model, that $^{12}$CO/$^{13}$CO is always equal to the $^{12}$C/$^{13}$C elemental ratio ($\sim$68). However, \citet{visser2009} were able, with their PDR model, to reproduce both higher and lower values with respect to the elemental ratio, as a function of H$_{2}$ column density. In fact, Visser et al. introduce Alfvén waves in their code, that have the effect of replacing the kinetic temperatures ($T_{\rm kin}$), in the rate equation of ion-neutral reactions, by an effective temperature that takes into account these non-thermal effects. They demonstrated that these effects can take place and modify the $^{12}$CO/$^{13}$CO ratio. 

Observationally, numerous studies of the $^{12}$C/$^{13}$C ratio have been conducted toward molecular clouds in the Galaxy, but most of them through observations of CO, H$_{2}$CO and HCO$^{+}$ that, as stated before, have the possible effects of isotopic-selective photodissociation and/or chemical fractionation (e.g. \citealt{wilson1994}; \citealt{langer1990}, and \citealt{langer1993}). To estimate the $^{12}$C/$^{13}$C elemental ratio across the Galaxy, \citet{milam2005} observed $^{12}$CN and $^{13}$CN toward Galactic molecular clouds. They found:
\begin{equation}
^{12}\textrm{C}/^{13}\textrm{C}=(6.01\pm1.19) \textrm{ kpc}^{-1}\times\textrm{D}_{\rm GC} +(12.28\pm9.33).
\label{milam}
\end{equation}
Moreover, they also derived a Galactocentric trend taking into account all together CN, CO and H$_{2}$CO observations:
\begin{equation}
[^{12}\textrm{C}/^{13}\textrm{C}]_{{\rm all\;molecules}}=(6.21\pm1.00) \textrm{ kpc}^{-1}\times\textrm{D}_{\rm GC} +(18.71\pm10.88),
\label{milam2}
\end{equation}
and then the average local present-day $^{12}$C/$^{13}$C is 68$\pm$15 (at a Sun distance $D_{\rm GC}$ = 7.9 kpc, as derived by \citealt{hunt2016} and \citealt{boehle2016}). In particular, the elemental \cratio\;ratio as a function of the Galactocentric distance is determined by stellar nucleosynthesis processes. \citet{romano2017} have shown that the observed trends are in agreement with Galactic chemical evolution models that include primary formation (i.e. starting from the primordial H and He nuclei) of $^{12}$C in all stars and of $^{13}$C in intermediate-mass asymptotic giant branch stars.

More recently, \citet{yan2019} presented H$_{2}$CO and H$_{2}^{13}$CO observations towards a sample of 112 sources from which they evaluated a new \cratio\;trend as a function of D$_{\rm GC}$. They obtained the following linear fit: 
\begin{equation}
^{12}\textrm{C}/^{13}\textrm{C}=(5.08\pm1.10) \textrm{ kpc}^{-1}\times\textrm{D}_{\rm GC} +(11.86\pm6.60),
\label{yan}
\end{equation}
which is consistent within the error bars with that found by \citet{milam2005} (both equation \ref{milam} and \ref{milam2}). This indicates that even if H$_{2}$CO can be affected by chemical fractionation processes, either as a consequence of reactions \eqref{C-reaction1} and \eqref{C-reaction2} or of isotope-selective photodissociation effects (e.g. \citealt{visser2009}), the trend with the Galactocentric distance can still be disentangled from these effects.

\citet{daniel2013} performed a non-local-thermal-equilibrium analysis of HCN, HNC, CN, and their $^{13}$C-isotopologues towards the pre-stellar core B1b. They found HNC/HN$^{13}$C=20$^{\raisebox{1pt}{\footnotesize\rlap{\eqmakebox[subp][l]{+5}}}}_{\footnotesize\raisebox{1pt}{\eqmakebox[subp][l]{-4}}}$, HCN/H$^{13}$CN=30$^{\raisebox{1pt}{\footnotesize\rlap{\eqmakebox[subp][l]{+7}}}}_{\footnotesize\raisebox{1pt}{\eqmakebox[subp][l]{-4}}}$, and CN/$^{13}$CN=50$^{\raisebox{1pt}{\footnotesize\rlap{\eqmakebox[subp][l]{+19}}}}_{\footnotesize\raisebox{1pt}{\eqmakebox[subp][l]{-11}}}$.
 \citet{magalhaes2018} obtained, towards the starless core L1498, a HCN/H$^{13}$CN ratio of 45$\pm$3. These works show that nitrile-bearing species are enriched in $^{13}$C with respect to the local ISM value (\cratio\;= 68). 

Knowledge of the exact value of the $^{12}$C/$^{13}$C ratio is also important for deriving the $^{14}$N/$^{15}$N ratio for nitrile-bearing species since most studies use the so-called double isotope method (e.g \citealt{wampfler2014}, \citealt{zeng2017}, \citealt{colzi18a}, \citealt{colzi18b}). This is based on observations of optically thin species, with $^{13}$C substituted for $^{12}$C (e.g. H$^{13}$CN, HN$^{13}$C).
However, this method works only if the assumed \cratio\;ratio is comparable, within the errors, with the value given by the Galactocentric trend, or if it can be determined independently from observations or chemical models. In fact, $^{13}$C chemical fractionation may affect the abundances, and also the \nratio\;ratio, of nitrile-bearing species. This behaviour is typically not taken into account in chemical models inclusive of $^{15}$N-bearing species. Moreover, the observed \cratio\;abundance ratios listed above are not consistent with the values predicted by chemical models for species like CN, HCN, and HNC. As shown by \citet{roueff2015}, the dilution of $^{13}$C for nitriles and isonitriles is at most a factor 2 if derived from pure gas-phase chemical models with a fixed kinetic temperature of 10 K. However, these models do not include gas-grain interactions and introduce depletion effects by simply varying the elemental abundances of Carbon, Oxygen and Nitrogen. It is thus important to test and refine the predictions of these models by introducing time dependent depletion effects resulting from gas-grain interactions. \citet{roueff2015} also studied C-fractionation for C, CH, CO, and HCO$^{+}$ showing that the \cratio\;ratio is highly time-dependent. In fact, since the start of the simulation, $^{13}$C$^{+}$ forms $^{13}$CO through reaction \eqref{C-reaction1}. For a similar reason, CN is enriched in $^{13}$C with similar time-scales compared to $^{13}$C-enrichment in CO, due to the reaction:
\begin{equation}
\label{eq-CNfrac}
^{13}{\rm C}^{+} + {\rm CN} \rightleftharpoons \hspace{0.1cm}^{12}{\rm C}^{+} + \hspace{0.1cm}^{13}{\rm CN} + 31.1\hspace{0.1cm}{\rm K}.
\end{equation} 
Reaction \eqref{C-reaction2} becomes important once $^{13}$C$^{+}$ starts to deplete from the gas phase.
HCO$^{+}$ is then enriched in $^{13}$C as long as the $^{13}$CO remains in excess with respect to $^{12}$CO (i.e. until CO/$^{13}$CO comes back to the initial value of 68). The gas-phase chemical model of \citet{roueff2015} reached a steady state at about 10$^{7}$ yr for a density of 2$\times$10$^{4}$ cm$^{-3}$. At this time C, CH, HCN, HNC, and CN are depleted in $^{13}$C. This is related to reactions \eqref{C-reaction1} and \eqref{C-reaction2} that continue to be important and retain all the $^{13}$C in CO and HCO$^{+}$.

In this work we report a new detailed analysis of the \cratio\;ratio derived under different fixed physical conditions, and with the introduction of new low-temperature isotopic exchange reactions. We first describe the gas-grain chemical model and the network used (Sect.~\ref{model}). Then, we present the fiducial model that we have chosen to describe the main characteristics and new results about C-fractionation of this chemical model (Sect.~\ref{fiducial}). Finally, we present and discuss a parameter-space exploration for CN, HCN, and HNC. We show how the C-fractionation in some specific models varies with the density, the temperature and the cosmic-ray ionization rate. We also describe the link with the \nratio\;derived towards high-mass star-forming regions with the double-isotope method (Sect.~\ref{parspace}).

\section{Model}
 \label{model}

\subsection{Chemical model}

Our chemical code is based on the one described in \citet{sipila2015a}, recently updated to include several new chemical processes (\citealt{sipila2019a}). In short, the code solves rate equations for gas-phase and grain-surface chemistry, which are connected through adsorption and (non-)thermal desorption. A description of the basic processes including relevant equations can be found in \citet{sipila2015a} and are not reproduced here for brevity.

\subsection{Introduction of $^{13}$C-fractionation in the chemical model}
\label{frac-procedure}
In this paper we model the isotopic fractionation of carbon by introducing the isotope $^{13}\rm C$ to the KIDA gas-phase network (\citealt{wakelam2015}) using a procedure similar to our earlier approach to generating deuterated networks (\citealt{sipila2013}, \citealt{sipila2015b}).

As an example of the fractionation-generation procedure, consider the reaction
\begin{equation}
\label{eq:example1}
\rm H + CCN \longrightarrow C + HCN \, .
\end{equation}
Here, the $^{13}\rm C$ atom can be introduced in CCN in two positions that are considered equivalent. From a chemical standpoint, the position of the $^{13}\rm C$ atom is important and may impact the reactivity of the molecule, but here we make the simplifying assumption that the order of carbon atoms does not need to be tracked. 
We note however that significant differences in abundances have been observed for molecules where the $^{13}$C atom can be located in different positions. For example, \citet{sakai2010} observed the \emph{N}=1--0 lines of CCH and its $^{13}$C-isotopic species towards the dark cloud TMC-1 and the star-forming region L1527. They investigated the $^{13}$C-species abundances and the possible different formation pathways. They found a C$^{13}$CH/$^{13}$CCH ratio of 1.6$\pm$0.4 and 1.6$\pm$0.1 towards TMC-1 and L1527, respectively. They proposed that the reaction
 \begin{equation}
 \label{eq-sakai1}
{\rm CH}_{2} +{\rm C} \rightarrow {\rm C}_{2}{\rm H} + {\rm H} 
 \end{equation}
 could make a difference between the formation of the two $^{13}$C-isotopic species of C$_{2}$H. \citet{taniguchi2019} found similar results towards L1521B and L134N. However, testing this result is beyond the scope of this work. \\
According to our present approach, the following branches will be generated once $^{13}\rm C$ is substituted in reaction\,\eqref{eq:example1}:
\begin{align}
\rm H &+ \rm C^{13}CN            \Arrow^{1/2}  C + H^{13}CN \nonumber \\
\rm H &+ \rm C^{13}CN            \Arrow^{1/2}  {^{13}}C + HCN \nonumber \\
\rm H &+ \rm {^{13}}C^{13}CN  \Arrow^1  {^{13}}C + H^{13}CN \, , \nonumber
\end{align}
where the branching ratio is displayed above the arrow. Essentially, we calculate the probability of a given branch based on the positions that the $^{13}\rm C$ atom can occupy on the {\sl product} side of the reaction. The same rule is applied to most reactions, i.e., the inherent assumption is that the reactions proceed via full scrambling, which is not universally true (\citealt{sipila2019b}). Notable exceptions to the full-scrambling rule are charge-exchange reactions and proton-donation reactions that involve at least one carbon atom in each reactant and product. For these reactions we assume that carbon atoms cannot be interchanged in the reaction. Mass and/or energy corrections to the rate coefficients of the reactions due to isotopic effects are expected to be small, below the expected accuracy of the reaction rate coefficients, except for possible isotopic exchange reactions, as discussed in Sect.~\ref{exchangereactions}. The same fractionation procedure is applied to our grain-surface network (\citealt{sipila2019a}).

We do not consider fractionation for all carbon-containing reactions included in KIDA, or in our surface network, in order to maintain relative simplicity while still including the chemistry that is essential for the molecules typically observed and used to derive fractionation ratios.
First, we discard all reactions that contain molecules with more than five atoms. So, for example, cyanoacetylene (HC$_{3}$N) is included in our models while methanol (CH$_{3}$OH) is not. We discuss in Sect.~\ref{results} the results on diatomic and triatomic molecules, which should not be affected too much by this restriction. Second, we only perform the fractionation procedure for reactions that contain molecules with up to three carbon atoms, and so our final networks do not contain species like $\rm C_3{^{13}C}H$. With these restrictions our final networks (gas-phase and grain-surface) contain a combined total of $\sim$11500 reactions. Appendix \ref{compar-roueff} describes the possible issues in the building of the isotopic chemistry and compares the present method to that used in \citet{roueff2015}.

\subsection{Isotopic exchange reactions}
 \label{exchangereactions}

\begin{table*}
\begin{center}
\caption{Carbon isotopic exchange reactions used in this work. The top panel displays the reactions already used by \citet{roueff2015}, while the bottom panel displays the reactions added for the present study.}
\begin{tabular}{lclcc}
  \hline
  Label  & Reaction & $k_{\rm f}$ & $f(B, m)$\tablefootmark{a} & $\Delta E$\tablefootmark{b}\\ 
          & & (cm$^{3}$ s$^{-1}$) & & (K)  \\
  \hline
  (1) B & $^{13}$C$^{+}$ + CO $\rightleftharpoons$ $^{12}$C$^{+}$ + $^{13}$CO & 6.6$\times$10$^{-10}\times (\frac{T}{300})^{-0.45} \times$ &1 & 34.7 \\
& & $ \exp(-6.5/T) \times \frac{1}{1+\exp(-34.7/T)}$ & & \\
(2)\tablefootmark{c} A & $^{13}$CO + HCO$^{+}$ $\rightleftharpoons$ CO + H$^{13}$CO$^{+}$ & 2.6$\times$10$^{-10} \times (\frac{T}{300})^{-0.4}$ &1 & 17.4 \\
(3) B& $^{13}$C$^{+}$ + CN $\rightleftharpoons$ $^{12}$C$^{+}$ + $^{13}$CN & 3.82$\times$10$^{-9} \times $  &1 & 31.1 \\
& &$ (\frac{T}{300})^{-0.4}\times \frac{1}{1+\exp(-31.1/T)}$ & & \\
(4) B& $^{13}$C + CN $\rightleftharpoons$ $^{12}$C + $^{13}$CN & 3.0$\times$10$^{-10} \times \frac{1}{1+\exp(-31.1/T)}$ & 1 & 31.1 \\
(5) B& $^{13}$C + C$_{2}$ $\rightleftharpoons$ $^{12}$C + $^{13}$CC & 3.0$\times$10$^{-10} \times \frac{2}{2+\exp(-25.9/T)}$ & 2 & 25.9 \\
\hline
 (6) B& $^{13}$C$^{+}$ + C$_{2}$ $\rightleftharpoons$ $^{12}$C$^{+}$ + $^{13}$CC &1.86$\times$10$^{-09}\times \frac{2}{2+\exp(-25.9/T)}$ & 2 & 25.9\\
 (7) B&  $^{13}$C$^{+}$ + $^{13}$CC  $\rightleftharpoons$ $^{12}$C$^{+}$ + $^{13}$C$_{2}$ & 1.86$\times$10$^{-09}\times \frac{0.5}{0.5+\exp(-26.4/T)}$ & 0.5 & 26.4\\
 (8) B& $^{13}$C + $^{13}$CC $\rightleftharpoons$ $^{12}$C + $^{13}$C$_{2}$ & 3.0$\times$10$^{-10}\times \frac{0.5}{0.5+\exp(-26.4/T)}$ & 0.5 & 26.4\\
 (9) B& $^{13}$C$^{+}$ +CS $\rightleftharpoons$ $^{12}$C$^{+}$ + $^{13}$CS & 1.86$\times$10$^{-09}\times \frac{1}{1+\exp(-26.3/T)}$  & 1 & 26.3\\
 (10) B & $^{13}$C + C$_{3}$ $\rightleftharpoons$ $^{12}$C + $^{13}$CC$_{2}$ & 3.0$\times$10$^{-10}\times \frac{2}{2+\exp(-27/T)}$ & 2 & 27\\
  (11) B& $^{13}$C$^{+}$ + C$_{3}$ $\rightleftharpoons$ $^{12}$C$^{+}$ + $^{13}$CC$_{2}$ & 9.5$\times$10$^{-10}\times \frac{2}{2+\exp(-27/T)}$  & 2& 27\\
 \hline
  \normalsize
  \label{tab-reaccolzi}
  \end{tabular}
  \end{center}
\tablefoot{Type A reactions are direct reactions, while type B reactions are those involving adduct formation, without isomerization, as defined by \citet{roueff2015}.\\
\tablefoottext{a}{$f(B, m)$ is a probability factor that depends on the rotational constant, mass, and symmetry factors of the reactants and products. In reactions involving $^{13}$C, the mass ratio of the reactants and the products is close to unity. Then, $f(B, m)= q({\rm C})q({\rm D})$/$q({\rm A})q({\rm B})$, where A and B are the reactants, C and D the products, and $q(...)$ are the internal molecular partition functions.}
\tablefoottext{b}{$\Delta E$ is the exoergicity of the reaction.}
\tablefoottext{c}{The published exponent in the rate coefficient in the A\&A journal is a misprint.}
}
 \end{table*}

In this work we examine the behaviour of C-fractionation in a low-temperature environment ($T_{\rm gas}\leq$50 K). Isotopic exchange reactions are very important in these cold regions and could affect the behaviour of the \cratio\;ratio in different molecules. \citet{roueff2015} studied the \cratio\;ratio of some molecules, introducing some isotopic exchange reactions. For this work we have updated the list of exchange reactions, as shown in Table \ref{tab-reaccolzi}.
These reactions can occur in the absence of potential barriers and when no other exothermic product channel is available.

We now discuss the new suggested reactions shown in the bottom panel of Table \ref{tab-reaccolzi}. To quantify the efficiency of these possible exchange reactions, we have computed the variation of the involved ZPE and we have assumed that the reactions take place via an intermediate complex. 

Since no experimental data is available, we have assumed a pre-exponential factor. For ion-neutral reactions we could use the rate given by the Langevin formula:
\begin{equation}
\label{eq-langevinrate}
k_{\rm AB}=2.34\times10^{-09} q \biggl(\frac{\alpha}{\mu}\biggr)^{\frac{1}{2}}\quad{\rm cm}^{3} {\rm s}^{-1},
\end{equation}
where $\mu$ is the reduced mass of the reactants in atomic mass units (amu), $q$ is the electronic charge, and $\alpha$ is the polarizability of the neutral species in cubic angstroms. However, since this expression represents an upper limit for the reaction probability, we decided to guess the rate coefficients, setting them to a similar order of magnitude with previously known reaction.
Reactions involving the C$^{+}$ ion and a diatomic molecule (6,7 and 9 of Table~\ref{tab-reaccolzi}) are assumed to proceed at the Langevin rate. However, the C$^{+}$ + C$_{3}$ reaction could lead to different products, depending on the carbon position. We assume for it a rate of 9.5$\times$10$^{-10}$~cm$^{3}$~s$^{-1}$, about half of the Langevin rate (1.7$\times$10$^{-9}$~cm$^{3}$~s$^{-1}$). We have made this assumption since in this work we consider only one position ($^{13}$C$^{12}$C$_{2}$).
Exchange reactions may occur in neutral-neutral reactions. In absence of more detailed theoretical investigations, for reactions (8) and (10) we assumed the same pre-exponential factor as for reaction (5) in Table \ref{tab-reaccolzi}.
Moreover, for the new reactions studied in this work, the factor $f$ is near unity unless C$_{2}$ (or C$_{3}$) appears, in which case it is $\sim$2.0 if the symmetric molecule C$_{2}$ (or C$_{3}$) is a reactant, and $\sim$0.5 if C$_{2}$ is a product, according to the symmetry factor term present in the partition function.

The exponential term $\exp(-\Delta E/T)$, which is both present in the forward reaction rate coefficients (except for reaction \eqref{C-reaction2}) and in the reverse (endothermic) ones, requires the knowledge of the energy defect ($\Delta$ZPE) which is obtained from the difference of the zero-point energies of the products and the reactants. In Table~\ref{table-deltazpe}, we report the difference in zero-point energies between the more and less abundant isotopologues found in the literature for C$_{2}$, CS, and C$_{3}$.
The energies involved depend on the position of $^{13}$C, as indicated in Table \ref{table-deltazpe}.
As we do not track the position of $^{13}$C, we take the lowest value involved for the energy defect, which allows us to evaluate the importance of this mechanism at its minimum level. 
$^{13}$C-fractionation in C$_{3}$ could proceed further through reaction of $^{13}$C with $^{13}$CC$_{2}$ and other $^{13}$C substituents. We made some tests which showed that the effect of multiple $^{13}$C-fractionation on our results is negligible. We leave the full study for future considerations.

\begin{table}
\setlength{\tabcolsep}{5pt}
\caption{Values of the ZPEs for molecules related to the isotopic exchange reactions used in our model. The related references are also given in the fourth column.}
\centering
  \begin{tabular}{lccc}
  \hline
  Molecule   & ZPE & $\Delta$ZPE\tablefootmark{a} & Reference\\ 
           & (cm$^{-1}$)&  (K)  &            \\
  \hline
 C$_{2}$  & 924.13 & -- & (1) \\
 $^{13}$CC & 906.1 & 25.9 & (1)\\
 $^{13}$C$_{2}$ & 887.8 & 52.3 & (1) \\
  $^{13}$C$_{2}^{b}$ &  & 26.4 & (1)\\
  \hline
  CS & 640.9 & -- & (2)\\
  $^{13}$CS & 622.7 & 26.3 &(3)\\
  \hline
  C$_{3}$ & 1705.06 & -- & (4) \\
  $^{13}$CC$_{2}$ & 1686.44 & 27 & (4) \\
  C$^{13}$CC & 1675.54 & 43 & (4)\\
  $^{13}$CC$^{13}$C\tablefootmark{b} & 1667.74 & 27 & (4) \\
  $^{13}$C$^{13}$CC\tablefootmark{b} & 1656.70 & 43 & (4)\\
  $^{13}$C$_{3}$\tablefootmark{c} & 1637.79 & 27& (4) \\
  \hline
  \normalsize
  \label{table-deltazpe}
  \end{tabular}
  \tablebib{
(1)~\citet{zhang2011}; (2) \citet{bergeman1981}; (3) \citet{huber1979}; (4) \citet{schroder2016}.
}
\tablefoot{
 \tablefoottext{a}{It is equal to $\Delta E$ used in Table~\ref{tab-reaccolzi}.} \tablefoottext{b}{$\Delta$ZPE is derived with respect to the second molecular species of the group.}
 \tablefoottext{c}{$\Delta$ZPE is derived with respect to the fifth molecular species of the group.}
}
      \normalsize
\end{table}

\section{Results and Discussion}
 \label{results}

In this section we discuss the main results obtained with the simulation at different densities and temperatures. The main physical parameters that the chemical code requires are the total number density of H nuclei ($n_{\rm H}$\footnote{$n_{\rm H}$ = $n$(H) + 2$n({\rm H}_{2})\simeq 2n({\rm H}_{2})$ in dense molecular clouds like those simulated in this work.}), the dust temperature ($T_{\rm dust}$), the gas temperature ($T_{\rm gas}$), the cosmic-ray ionization rate ($\zeta$), the visual extinction ($A_{\rm V}$), the grain albedo ($\omega$), the grain radius ($a_{\rm g}$), the grain material density ($\rho_{\rm g}$). The ratio between the diffuse and the binding energy of a species on dust grains ($\epsilon$), and the dust-to-gas mass ratio ($R_{\rm g}$). Apart from densities and temperatures, we have fixed all of the other initial parameters as described in Table \ref{table-physpar}, except in Sect.~\ref{effectCR} where we have performed an analysis varying the cosmic-ray ionization rate.

\begin{table}
\caption{Values of the physical parameters fixed in each model.}
  \begin{tabular}{lc}
  \hline
  Parameter   & Value\\ 
  \hline
 $\zeta$  & 1.3$\times$10$^{-17}$ s$^{-1}$\\
  $A_{\rm V}$   & 10 mag \\
  $\omega$& 0.6\\
  $a_{\rm g}$ & 10$^{-5}$ cm\\
  $\rho_{\rm g}$& 3 g cm$^{-3}$ \\
  $\epsilon= E_{\rm diff}/E_{\rm b}$ &0.6 \\
  $R_{\rm d}$= dust-to-gas mass ratio & 0.01 \\
  \hline
  \normalsize
  \label{table-physpar}
  \end{tabular}
\centering
\end{table}

We assumed that the gas is initially atomic except for hydrogen which is in molecular form. The adopted initial abundances are presented in Table \ref{table-initab} and are used for all of the models presented in this work, assuming an initial \cratio\;ratio of 68. 

\begin{table}
\setlength{\tabcolsep}{6pt}
\caption{Initial abundances with respect to $n_{\rm H}$. Adapted from \citet{semenov2010}.}
\centering
  \begin{tabular}{cc}
  \hline
  Species   & Initial abundance\\ 
  \hline
  H$_{2}$   & 0.5\\
  He    & 9.00$\times$10$^{-2}$\\
  C$^{+}$& 1.20$\times$10$^{-4}$\\
  $^{13}$C$^{+}$ & 1.76$\times$10$^{-6}$\\
  N    & 7.60$\times$10$^{-5}$ \\
  O & 2.56$\times$10$^{-4}$\\
   S$^{+}$   & 8.00$\times$10$^{-8}$\\
   Si$^{+}$        &8.00$\times$10$^{-9}$ \\
   Na$^{+}$         &2.00$\times$10$^{-9}$ \\
  Mg$^{+}$         & 7.00$\times$10$^{-9}$\\
Fe$^{+}$     & 3.00$\times$10$^{-9}$\\
P$^{+}$      &2.00$\times$10$^{-10}$ \\
   Cl$^{+}$      & 1.00$\times$10$^{-9}$\\
 F    & 2.00$\times$10$^{-9}$\\
      \hline
  \normalsize
  \label{table-initab}
  \end{tabular}
 \tablefoot{The initial $^{13}$C abundance is given by \cratio =68.
 }
\end{table}

 \subsection{The fiducial model}
 \label{fiducial}
 First of all, we analysed the behaviour of the abundances and \cratio\;ratios of different species for a particular model with a fixed temperature and density. We chose as our fiducial model the one with $T_{\rm gas}$=10 K and $n_{\rm H}$ = 2$\times$10$^{4}$ cm$^{-3}$. Note that we always assume that the gas and dust are thermally coupled, so that $T_{\rm gas}$=$T_{\rm dust}$.
\subsubsection{Gas-phase model}

\begin{figure*}
\centering
\includegraphics[width=45pc]{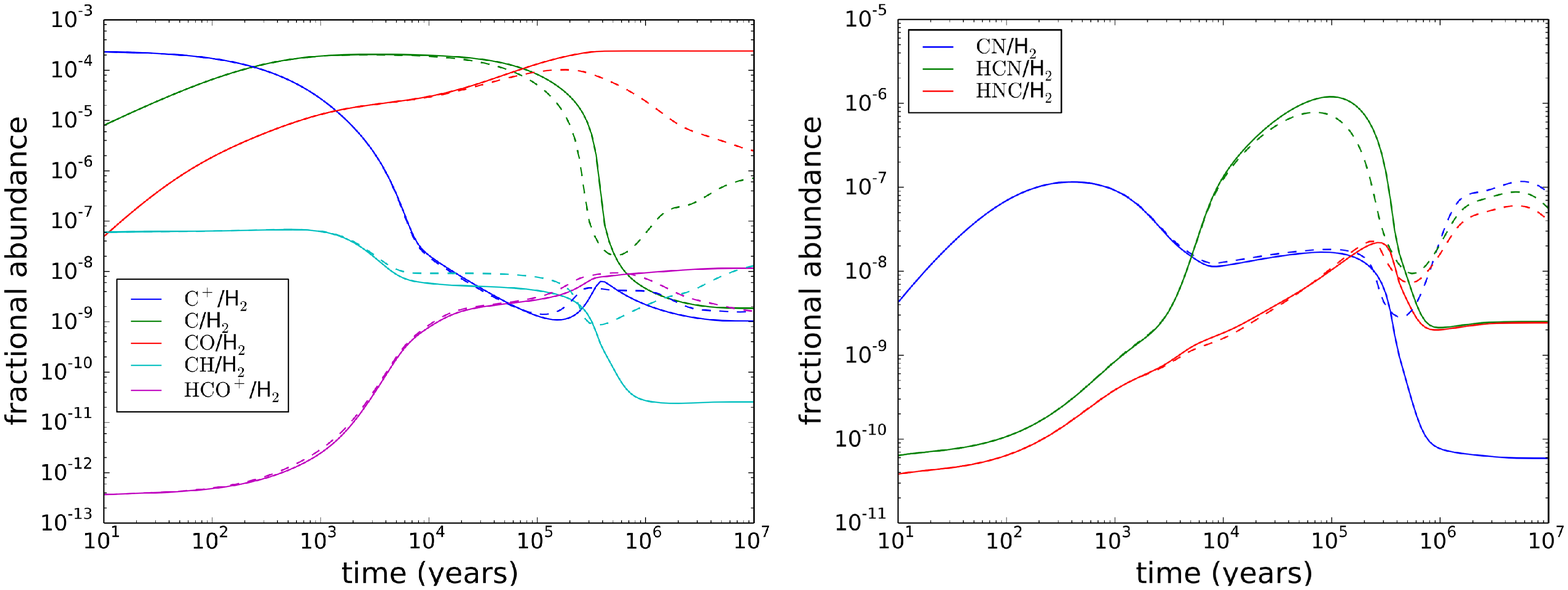}
\caption{Time evolution of C$^{+}$, C, CO, CH, and HCO$^{+}$ (\emph{left panel}) and of CN, HCN, and HNC (\emph{right panel}) abundances with respect to H$_{2}$ for the fiducial model with only gas-phase chemistry at work (solid lines) and with both gas-phase and grain-surface chemistry at work (dashed lines).}
\label{fig-abu-gasgrain-over}
\end{figure*}
\begin{figure*}
\centering
\includegraphics[width=45pc]{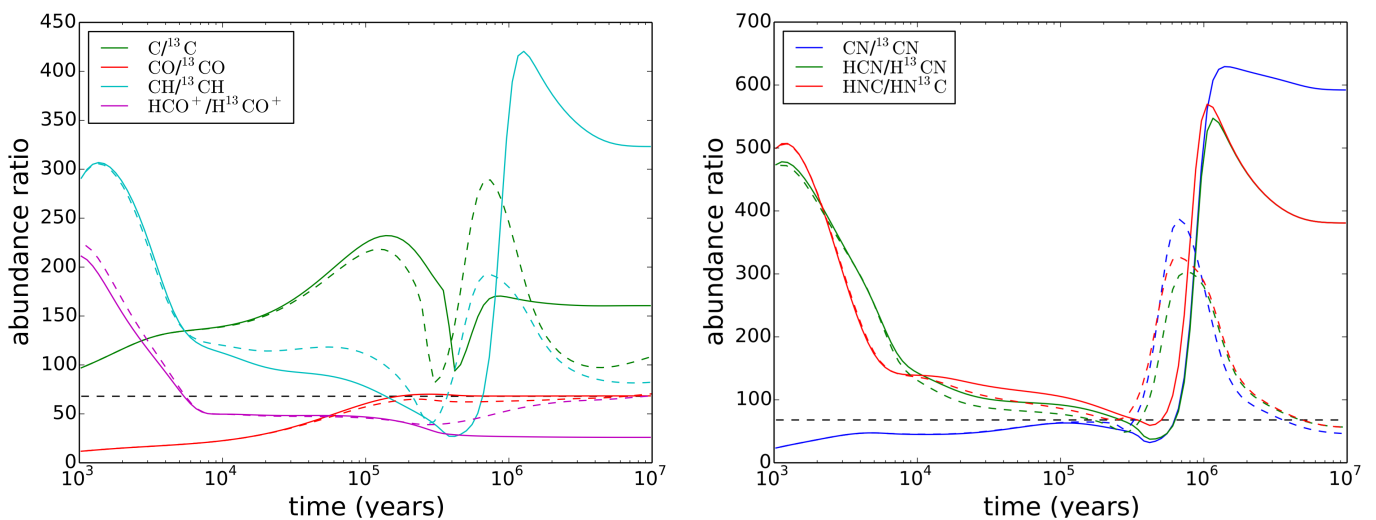}
\caption{Time evolution of the \cratio\;ratio for C, CO, CH and HCO$^{+}$ (\emph{left panel}), and for CN, HCN, and HNC (\emph{right panel}) for the fiducial model with only gas-phase chemistry at work (solid lines) and with both gas-phase and grain-surface chemistry at work (dashed lines). In both panels the black horizontal dashed line represents the initial \cratio\;ratio of 68.}
\label{fig-ratio-gasgrain-over}
\end{figure*}

 Fig.~\ref{fig-abu-gasgrain-over} shows the time-dependence of C$^{+}$, C, CO, CH, HCO$^{+}$, CN, HCN, and HNC abundances, for a model with only gas-phase chemistry at work (solid lines). Here we simulate the formation of H$_{2}$ in the grain-surface H + H association by constructing a dummy gas-phase reaction. The rate coefficient for this reaction is derived from the grain physical parameters reported in Table~\ref{table-physpar}, i.e. $\frac{1}{2} s \frac{3}{4}\frac{1.4 m_{\rm H} R_{\rm d}}{\rho a_{\rm g}}  $v$_{\rm H} n_{\rm H}$, where $s$ is the hydrogen sticking coefficient, $m_{\rm H}$ is the mass of the hydrogen atom, and v$_{\rm H}$ is the thermal speed of hydrogen (\citealt{lepetit2002}).
 As in model (a) in \citet{roueff2015}, which corresponds to the same temperature, density and cosmic-ray ionization rate, the steady-state is reached at a few million years. At the beginning, all the carbon is in ionized form, and after 10$^{2}$ yr it is converted into atomic carbon, which is later transformed into CO (in a time-scale of $\sim$10$^{5}$ yr). Meanwhile, CH follows the behaviour of atomic carbon, and HCO$^{+}$ that of CO, as expected. 
In Fig.~\ref{fig-ratio-gasgrain-over} the related \cratio\;ratios are shown with solid lines. As already explained in Sect.~\ref{intro}, the \cratio\;ratios of these molecules are mainly governed by the isotopic exchange reactions (1), (2), and (3) shown in Table \ref{tab-reaccolzi}. However, this model already shows differences between the results of \citet{roueff2015} and this work. In fact, there is a range of time in which the \cratio\;ratios for nitrile-bearing species tend to be similar and lower than 68. These values for HCN and HNC were different and always higher than 68 in previous chemical models that simulated similar physical conditions (e.g. \citealt{roueff2015}). As we will discuss in Sect.~\ref{importanceC3}, this behaviour is due to the introduction of the carbon isotopic exchange reaction of C$_{3}$.

\subsubsection{Gas-grain model}
\label{sec-gas+grain}

The time-dependence of C$^{+}$, C, CO, CH, HCO$^{+}$, CN, HCN, and HNC abundances, for the complete model, with both gas-phase and grain surface reactions, is shown in Fig.~\ref{fig-abu-gasgrain-over} with dashed lines. In addition to what happens with an only gas-phase model, after 2$\times$10$^{5}$ yr CO starts to freeze-out on grain surfaces and the abundance in the gas-phase drops. 
Moreover, when most of the CO is depleted, the abundance of He$^{+}$, which is produced at a constant rate by cosmic-ray-induced ionization, increases as CO is one of its main destruction partner. He$^{+}$ continues to react with the remaining CO in gas-phase, increasing the abundance of C$^{+}$. Moreover, as depletion occurs, C$^{+}$ mainly reacts with H$_{2}$ than with O-bearing species. This increases the CH abundance, and as a consequence the abundance of related molecules, like C$_{2}$, C$_{3}$, CN, HCN, and HNC, increases as well (Fig.~\ref{fig-abu-gasgrain-over} and left panel of Fig.~\ref{fig-C3-behaviour}). This behaviour was already predicted by \citet{ruffle1997}, and carbon-chains such as HC$_{3}$N can be used as depletion indicators in late-type chemistry (after CO freeze-out).

Dashed lines in Fig.~\ref{fig-ratio-gasgrain-over} display the time dependence of the \cratio\;ratios of the same molecules discussed above for the gas-phase model. For t$<$10$^{5}$ yr, the inclusion of gas-grain interactions does not affect the isotopic ratio of C-bearing molecules. In fact, adsorption rates depend on v$_{\rm i}$, the thermal speed of species $i$, that is inversely proportional to its mass. Since typical differences in mass between $^{12}$C-containing species and $^{13}$C-containing species are less than a few percent, they do not have a significant effect. CO shows a $^{13}$C-enhancement until the last million years of the time interval covered by the simulation.
The $^{12}$C/$^{13}$C ratios of C, CH, CN, HCN and HNC peak at around 10$^{6}$ yr. The peak appears because of the interaction between gas and grains. In fact, during the fast early time chemistry most of the atomic carbon is transformed into CO. During this period, most of the $^{13}$C is in principally in CO and, to a less extent, in HCO$^{+}$, and is unavailable for other species. Then, after 10$^{6}$ yr the $^{12}$C/$^{13}$C ratios of the molecules listed above tend to decrease again because of the CO freeze-out that drives again atomic C to be the main reservoir of gas-phase carbon.
Moreover, before 10$^{6}$ yr there is a short interval of time where the $^{12}$C/$^{13}$C ratios of C, CH, CN, HCN and HNC tend to decrease, down to values lower than 68 (only for CH, CN and HCN), and after that time-scale the ratios tend to increase again.
We discuss below why this feature in the time dependence of $^{12}$C/$^{13}$C~ratio is present, and its consequences at different densities and temperatures.

In Appendix \ref{compar-furu} we compare the results obtained with our gas-grain chemistry with the ones obtained by \citet{furuya2011}. The results of the two chemical models are in agreement, and the differences are mainly due to the new low-temperature isotopic exchange reactions introduced in this work.

\subsubsection{The importance of the possible C$_{3}$ isotopic exchange reaction}
\label{importanceC3}

\begin{figure*}
\centering
\includegraphics[width=45pc]{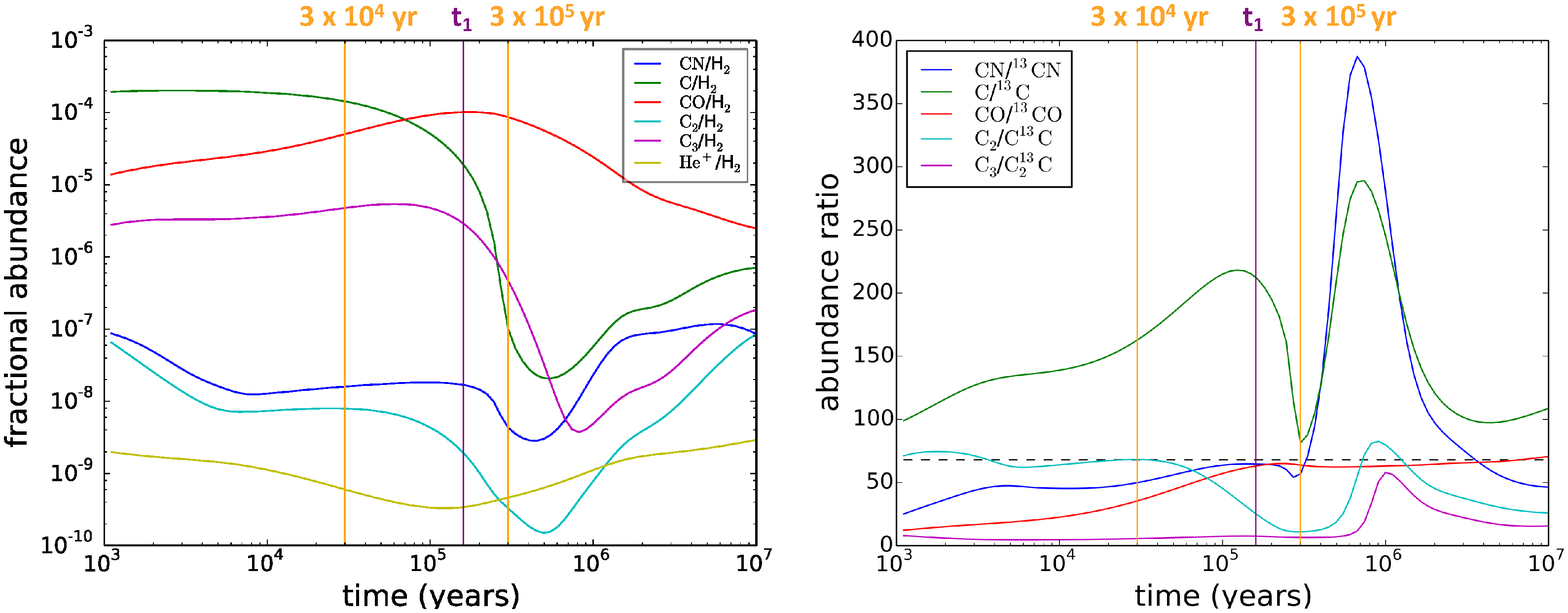}
\caption{\emph{Left panel:} Time evolution of CN, C, CO, C$_{2}$, C$_{3}$, and He$^{+}$ abundances with respect to H$_{2}$ for the fiducial model. \emph{Right panel:} Time evolution of the \cratio\;for CN, C, CO, C$_{2}$, and C$_{3}$ for the fiducial model. The black horizontal dashed line represents the initial \cratio\;ratio of 68. In both panels, the vertical purple solid line represents the "early chemistry time" as defined in the text, while the two vertical orange solid lines represent the two times (3$\times$10$^{4}$ yr, and 3$\times$10$^{5}$ yr) that we have analysed and discussed in the text.}
\label{fig-C3-behaviour}
\end{figure*}
\begin{figure*}
\centering
\includegraphics[width=45pc]{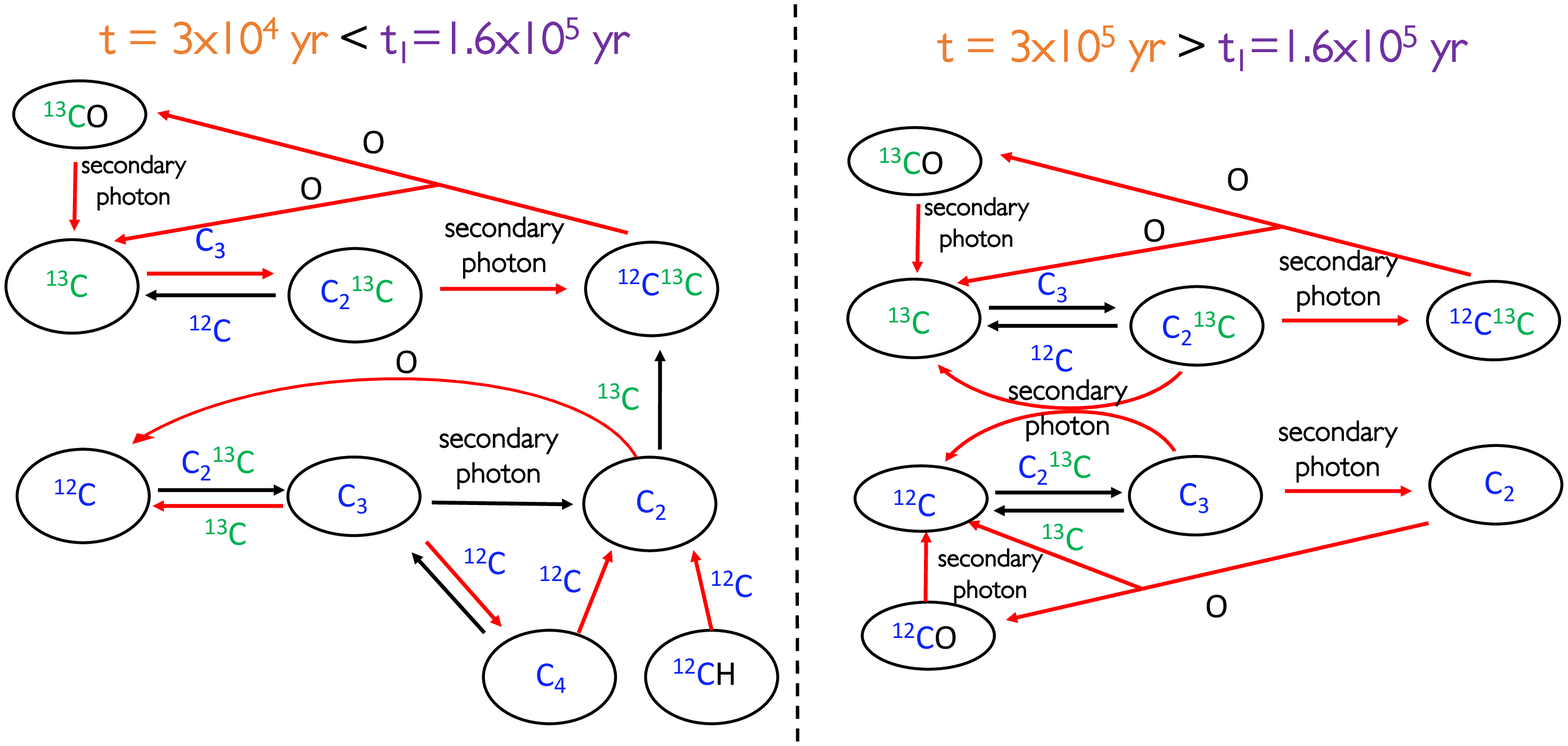}
\caption{Chemical pathways that distribute the two carbon isotopes in atomic carbon, C$_{2}$ and C$_{3}$ at 3$\times$10$^{4}$ yr (\emph{left panel}) and 3$\times$10$^{5}$ yr (\emph{right panel}), for the fiducial model. Main creation and destruction reactions are highlighted in red, $^{12}$C is represented in blue and $^{13}$C is represented in green.}
\label{fig-13C-reactions}
\end{figure*}
Observations of atomic carbon, and simple molecules containing more than one carbon atom, are important to put constraints on the processes that form larger molecules. The linear molecule C$_{3}$ is one of these species. It was observed for the first time in the ISM by \citet{haffner1995}, who reported a tentative detection, and clearly by \citet{maier2001}. Later \citet{roueff2002} studied this molecule towards the diffuse molecular cloud HD 210121 where it forms mainly from the recombination of C$_{3}$H$^{+}$, and is destroyed by photodissociation. They found an abundance relative to H$_{2}$ of 6.75$\times$10$^{-9}$. Other detections were made later towards stars surrounded by molecular clouds and in translucent sight lines (\citealt{galazutdinov2002}, \citealt{adamkovics2003} and \citealt{oka2003}).
Moreover, \citet{wakelam2009} pointed out the possible occurrence of radiative association of C$_{3}$ and C to form C$_{4}$ and the formation channel of CO through C$_{4}$ + O. They also pointed out that the reaction C + C$_{5}$ can be important for producing more C$_{3}$ and CO. 
Finally, \citet{mookerjea2012} and \citet{mookerjea2014} observed and modelled the abundance of C$_{3}$ towards envelopes of high-mass star-forming regions. \citet{mookerjea2012} found an abundance of C$_{3}$ of (6.3$\pm$1.3)$\times$10$^{-10}$ in the envelope towards DR1(OH), that they were able to reproduce with a chemical model with $n_{\rm H_{2}}$=5$\times$10$^{6}$ cm$^{-3}$ and a temperature of 30 K. \citet{mookerjea2014} observed along the line of sight of the UC HII region W51e2, with \emph{Herschel}, detecting an absorption feature probably tracing a cold external envelope. 

The carbon isotopic exchange reaction involving C$_{3}$:
\begin{equation}
\label{reac-C3}
 ^{13}{\rm C} + {\rm C}_{3} \rightleftharpoons \hspace{0.1cm}^{12}{\rm C} + \hspace{0.1cm}^{13}{\rm CC}_{2} + 27\hspace{0.1cm}{\rm K},
\end{equation}
was already emphasized in the discussion of the results displayed in Fig.~\ref{fig-ratio-gasgrain-over}.
In this section, we investigate the contribution of this species to the \cratio\;ratio of different molecules. For this, we studied in detail the main reactions that form or destroy C$_{3}$ at two precise times in the fiducial model: 3$\times$10$^{4}$ yr and 3$\times$10$^{5}$ yr. These two times are earlier and later than the "early chemistry time" ($t_{\rm 1}$), that we have defined as the time at which the abundance of atomic carbon drops by one order magnitude while it is transformed into CO. 
In the fiducial model $t_{\rm 1}$=1.6$\times$10$^{5}$ yr, and it is shown in Fig.~\ref{fig-C3-behaviour} as the vertical purple line.
Fig.~\ref{fig-C3-behaviour} displays the abundances (left panel) and the \cratio\;ratio (right panel) for CN, C, CO, C$_{2}$, and C$_{3}$. We note that the abundance of C$_{3}$ is two orders of magnitude higher than that of CN until the late-chemistry time, when CO starts to freeze-out onto grain surfaces ($\sim$2--4$\times$10$^{5}$ yr).
The main reactions for the two times are summarised in Fig.~\ref{fig-13C-reactions} and are explained below in detail. Moreover, the \cratio\;ratio for some important molecules at the three times (3$\times$10$^{4}$ yr, 1.6$\times$10$^{5}$ yr and 3$\times$10$^{5}$ yr) are summarised in Table~\ref{table-fiducial-cratio}.

\begin{itemize}
\item For $t<t_{\rm 1}$, $^{13}$CC$_{2}$ is enriched in $^{13}$C, with respect to C$_{3}$, thanks to the forward reaction \eqref{reac-C3}. As a consequence, the C$_{3}$/$^{13}$CC$_{2}$ ratio is lower than the elemental initial value of 68, and the atomic \cratio\;is higher than 68. Even though reaction \eqref{eq-CNfrac} has a similar exothermicity, reaction \eqref{reac-C3} is more efficient since C$_{3}$/H$_{2}>$ CN/H$_{2}$. This means that the C$_{3}$/$^{13}$CC$_{2}$ ratio is lower than the CN/$^{13}$CN ratio and it stays low for a longer time than that of CN (CN/$^{13}$CN=68 at $t_{\rm 1}$). Moreover, most of C$_{2}$ is formed from atomic $^{12}$C through the cycle:
\begin{equation}
\label{eq-12Ccycle}
{\rm C}_{2} \xrightarrow{{\rm O}} {\rm C} \xrightarrow{{\rm C}_{4},\;{\rm CH}} {\rm C}_{2}.
\end{equation}
Conversely, C$^{13}$C is mainly formed from $^{13}$CC$_{2}$ through secondary UV photon reactions\footnote{The inner parts of a dense core are shielded from external UV photons, which can in the inner regions be created only through cosmic-ray-induced H$_{2}$ electronic excitation.}. As a consequence, C$_{2}$/$^{13}$CC$\simeq$68 since it reaches an equilibrium between the efficient formation of C$_{2}$ from the very abundant $^{12}$C and the main formation of C$^{13}$C from $^{13}$CC$_{2}$.

\item For $t>t_{\rm 1}$, the cycle of reactions \eqref{eq-12Ccycle} is not efficient any more since most of the atomic carbon has been transformed into CO. Thus, C$_{2}$ is mainly formed from C$_{3}$ reacting with secondary photons. C$_{2}$ then reacts with the remaining O, releasing atomic carbon. At this time, the cycle of reactions described above is the same for $^{12}$C and $^{13}$C-containing species (red reactions in the right panel of Fig.~\ref{fig-13C-reactions}). Since at this time-scale the isotopic exchange reaction of C$_{3}$ is not efficient any more, the atomic \cratio\;ratio approaches the low values of C$_{3}$. As a consequence the \cratio\;ratio of carbon-chains and nitrile-bearing species, that are produced starting from atomic carbon, is low as well. This behaviour remains until the abundance of C$_{3}$ drops (when CO starts to freeze-out on dust grains) and the \cratio\;ratios of the other molecular species rise again. When this happens, the main sink of $^{13}$C is CO and, to a small extent, HCO$^{+}$, and $^{13}$C is diluted in all the other molecular species.

\end{itemize}

We would like to point out that in our chemical network we did not introduce $^{13}$C in reactions containing C$_{4}$, as explained in Sect.~\ref{frac-procedure}, and some biases on C-fractionation of C$_{3}$ could arise because of this assumption. However, we found that the C$_{4}$ + O reaction is not efficient for the formation of C$_{3}$ in our chemical network. Moreover, the radiative association of C$_{3}$ and C giving C$_{4}$ and the C$_{4}$ + C reaction forming C$_{3}$ back, are much less efficient than the isotopic exchange reactions involving C$_{3}$ with our assumed reaction rate coefficients.

Recently, \citet{giesen2020} have reported the first detection of the $^{13}$C-isotopologues of C$_{3}$, $^{13}$CCC and C$^{13}$CC, towards the massive star-forming region SgrB2(M), near the Galactic Center. They derived an average \cratio\;abundance ratio of 20.5$\pm$4.2, in agreement with the value of 20 as derived from the Galactocentric trend by \citet{milam2005} for CN, CO, and H$_{2}$CO, taking into account the three molecules (equation \ref{milam2}).
Thus, it seems that no $^{13}$C-fractionation for C$_{3}$ is detected in SgrB2, contrary to our fiducial model predictions. However, this chemical model is not appropriate to model the physical conditions and the chemistry towards this source. Furthermore, other observational studies towards the Galactic Center are needed to improve the reliability of the \cratio\;estimated from the Galactocentric trend.
Finally, the authors found a N($^{13}$CCC)/N(C$^{13}$CC) ratio of 1.2$\pm$0.1, different from the statistically expected value\footnote{Assuming a statistical distribution of $^{13}$C in C$_{3}$, there are two options to place it at the ends of the carbon chain and only one at the center of the molecule.} (2). This discrepancy could be explained by a difference of 16 K in zero-point energy between the two species with respect to the main species C$_{3}$, as shown in Table~\ref{table-deltazpe}. The different position of the $^{13}$C in carbon chains would probably lead to a higher abundance of C$^{13}$CC with respect to $^{13}$CCC for 27~K~$< T <$~43 K because of the reaction C$_{3}$~+~$^{13}$C~$\rightarrow$~C$^{13}$CC~+~C~+~43 K. Another difference is that the symmetry factor $f$ for the reaction C$_{3}$~+~$^{13}$C~$\rightarrow$~C$^{13}$CC~+~C is~1, while is 2 for the reaction C$_{3}$~+~$^{13}$C~$\rightarrow$~$^{13}$CCC~+~C. We plan to upgrade our chemical network to track the position of $^{13}$C in molecules and hence to study the possible roles of the different ZPEs.

\begin{table}
\setlength{\tabcolsep}{5pt}
\caption{\cratio\;ratios for different molecules (C$_{3}$, C$_{2}$, C, CO, CN, HCN, and HNC) for three different times in the fiducial model: the "early chemistry time" (second row), $t_{1}$, as defined in the text, and the times before and after $t_{1}$ that we have discussed in the text (first and third row, respectively).}
  \begin{tabular}{lccccccc}
  \hline
  Time   & \multicolumn{7}{c}{\cratio} \\ 
          & C$_{3}$   & C$_{2}$ &C & CO & CN &HCN & HNC       \\
\hline
 3$\times$10$^{4}$ yr  &6& 68  & 163  & 35&50 &88 &106\\
  $t_{1}$=1.6$\times$10$^{5}$ yr   & 7 & 24 & 211  & 63  & 65 & 68 & 76\\
  3$\times$10$^{5}$ yr    &6 &11 & 81& 64&  57  & 50 & 79\\
  \hline
  \normalsize
  \label{table-fiducial-cratio}
  \end{tabular}
  \centering
\end{table}

\subsection{Parameter-space exploration}
\label{parspace}

\begin{figure*}
\centering
\includegraphics[width=38pc]{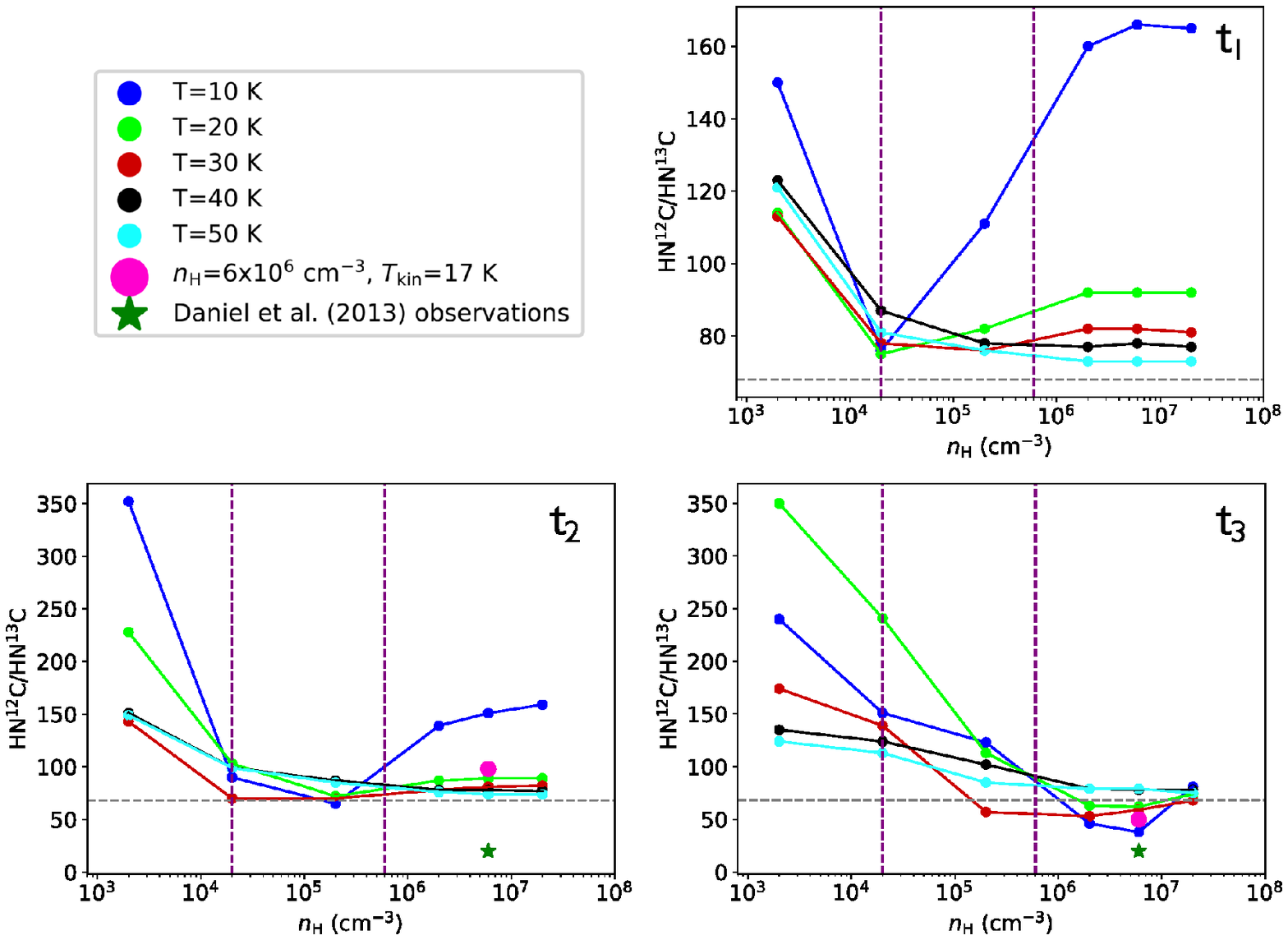}
\caption{HN$^{12}$C/HN$^{13}$C ratio as a function of $n_{\rm H}$, for different temperatures, at $t_{\rm 1}$ (\emph{top right panel}), $t_{\rm 2}$ (\emph{bottom left panel}) and $t_{\rm 3}$ (\emph{bottom right panel}). The large pink circles in the bottom panels represent the result of the model that simulates the physical conditions in the center of the pre-stellar core B1b observed by \citet{daniel2013}. The HN$^{12}$C/HN$^{13}$C ratio observed by \citet{daniel2013} is represented by the dark green star. Note that the error bar is within the symbol. In all the panels the two vertical purple dashed lines represent the range of densities of the sample of high-mass star-forming region described in Sect.~\ref{high-mass}, and the black horizontal dashed lines represent the initial \cratio\;ratio of 68.}
\label{fig-HNC-parspaceexplor}
\end{figure*}

\begin{figure*}
\centering
\includegraphics[width=38pc]{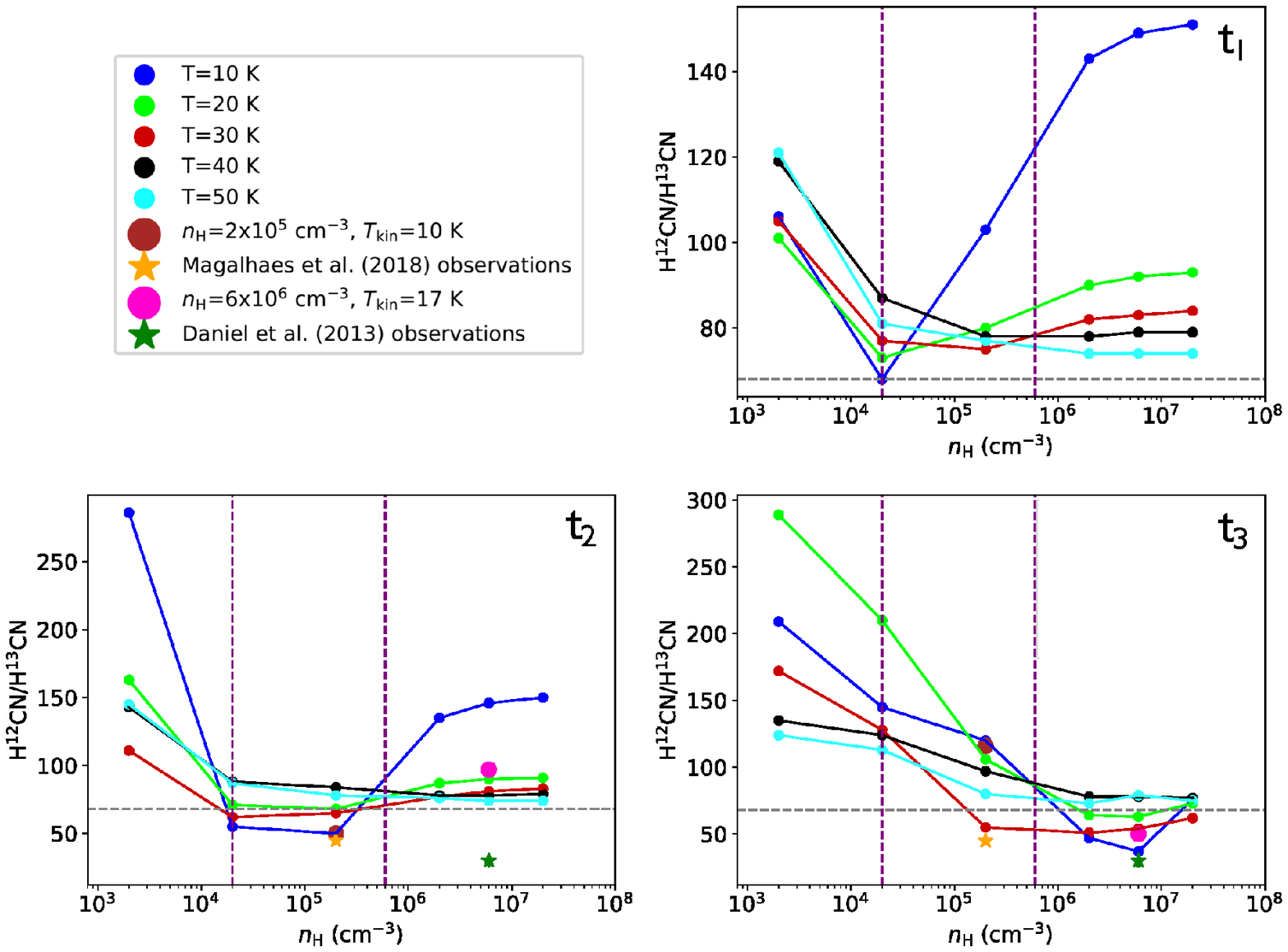}
\caption{H$^{12}$CN/H$^{13}$CN ratio as a function of $n_{\rm H}$, for different temperatures, at $t_{\rm 1}$ (\emph{top right panel}), $t_{\rm 2}$ (\emph{bottom left panel}), and $t_{\rm 3}$ (\emph{bottom right panel}). The large pink and brown circles in the bottom panels represent the result of the model that simulates the physical conditions in the center of the pre-stellar cores B1b observed by \citet{daniel2013} and L1498 observed by \citet{magalhaes2018}, respectively. The H$^{12}$CN/H$^{13}$CN ratios observed by \citet{daniel2013} and \citet{magalhaes2018} are represented by the dark green and orange stars, respectively. Note that the error bars are within the symbols. In all the panels the two vertical purple dashed lines represent the range of densities of the sample of high-mass star-forming region described in Sect.~\ref{high-mass}, and the black horizontal dashed lines represent the initial \cratio\;ratio of 68.}
\label{fig-HCN-parspaceexplor}
\end{figure*}

\begin{figure*}
\centering
\includegraphics[width=38pc]{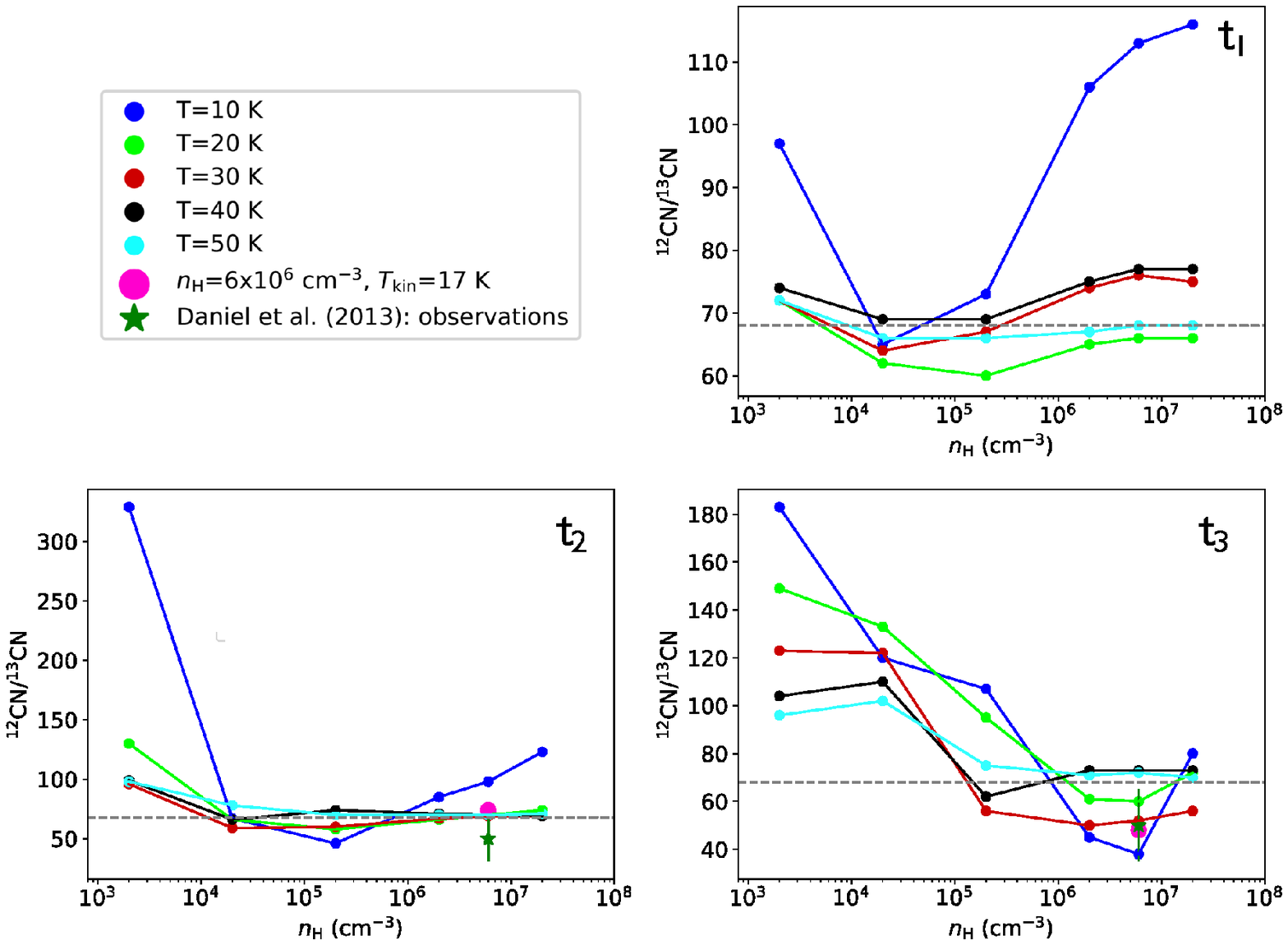}
\caption{$^{12}$CN/$^{13}$CN ratio as a function of $n_{\rm H}$, for different temperatures, at $t_{\rm 1}$ (\emph{top right panel}), $t_{\rm 2}$ (\emph{bottom left panel}), and $t_{\rm 3}$ (\emph{bottom right panel}). The large pink circles in the bottom panels represent the result of the model that simulates the physical conditions in the center of the pre-stellar core B1b observed by \citet{daniel2013}. The $^{12}$CN/$^{13}$CN ratio observed by \citet{daniel2013} is represented by the green star. Note that the error bar is within the symbol. In all the panels the black horizontal dashed line represents the initial \cratio\;ratio of 68.}
\label{fig-CN-parspaceexplor}
\end{figure*}

In this section, we focus the analysis on the nitrile-bearing species CN, HCN, and HNC in order to evaluate how reliable the estimates of the \nratio\;ratio based on \cratio = 68 (\citealt{colzi18a}; \citealt{colzi18b}) are.

We analysed the \cratio\;ratio for different temperatures and densities in three well defined fixed times: the "early chemistry" time $t_{\rm 1}$, $t_{\rm 2}$=2$\times t_{\rm 1}$ and $t_{\rm 3}$=10$\times t_{\rm 1}$. The densities we analysed are between 2$\times$10$^{3}$ cm$^{-3}$ and 2$\times$10$^{7}$ cm$^{-3}$, and the temperatures are 10, 20, 30, 40 and 50 K.
Figs.~\ref{fig-HNC-parspaceexplor}, \ref{fig-HCN-parspaceexplor}, and \ref{fig-CN-parspaceexplor} show the behaviour of HN$^{12}$C/HN$^{13}$C, H$^{12}$CN/H$^{13}$CN, and $^{12}$CN/$^{13}$CN, respectively, as a function of $n_{\rm H}$ and for different temperatures. In particular, it can be noted that for $t_{\rm 1}$ and $t_{\rm 2}$ and for densities higher than 10$^{6}$~cm$^{-3}$, the \cratio\;ratios tend to be higher for 10 K with respect to higher temperatures. This is because the forward reaction \eqref{C-reaction2} is efficient until T$<$17.4 K, together with the low-temperature isotopic exchange reaction for C$_{3}$. Then, most of the $^{13}$C is in CO, HCO$^{+}$, and in C$_{3}$.
It should be noted that this trend is also different depending on the density. This is because of the definition of $t_{\rm 1}$: for a density of 10$^{6}$ cm$^{-3}$ $t_{1}$ is the time just before reaching the dip of carbon isotopic ratios that directly follows C$_{3}$, while for a density of 10$^{5}$ cm$^{-3}$ $t_{1}$ is the time when the dip in \cratio\;is present and follows that of C$_{3}$. The higher the density, the later the dip of lower \cratio\; with respect to $t_{\rm 1}$ when most of the atomic C is in the form of CO. The most probable explanation, following also what was  described in Sect.~\ref{importanceC3}, is that there is a competition between the O atoms transformed into CO and frozen out on grains, and those that are still available to take part in the reactions shown in the right panel of Fig.~\ref{fig-13C-reactions}.

Overall, for densities of $\sim$10$^{3}$--10$^{4}$ cm$^{-3}$ the \cratio\;ratio is almost always higher than the canonical value of 68. Moreover, for a density $\simeq$10$^{5}$ cm$^{-3}$, the \cratio\;ratio is consistent with 68 within a factor two, except for HCN and HNC for temperatures below 20 K at $t_{\rm 1}$ and $t_{\rm 3}$.

We also performed the same parameter-space exploration for CO, HCO$^{+}$, and H$_{2}$CO, to show the predicted \cratio\;ratios also for these molecular species (see Appendix \ref{parspace-other}).

\subsubsection{Low-mass star-forming regions}
\begin{figure*}
\centering
\includegraphics[width=38pc]{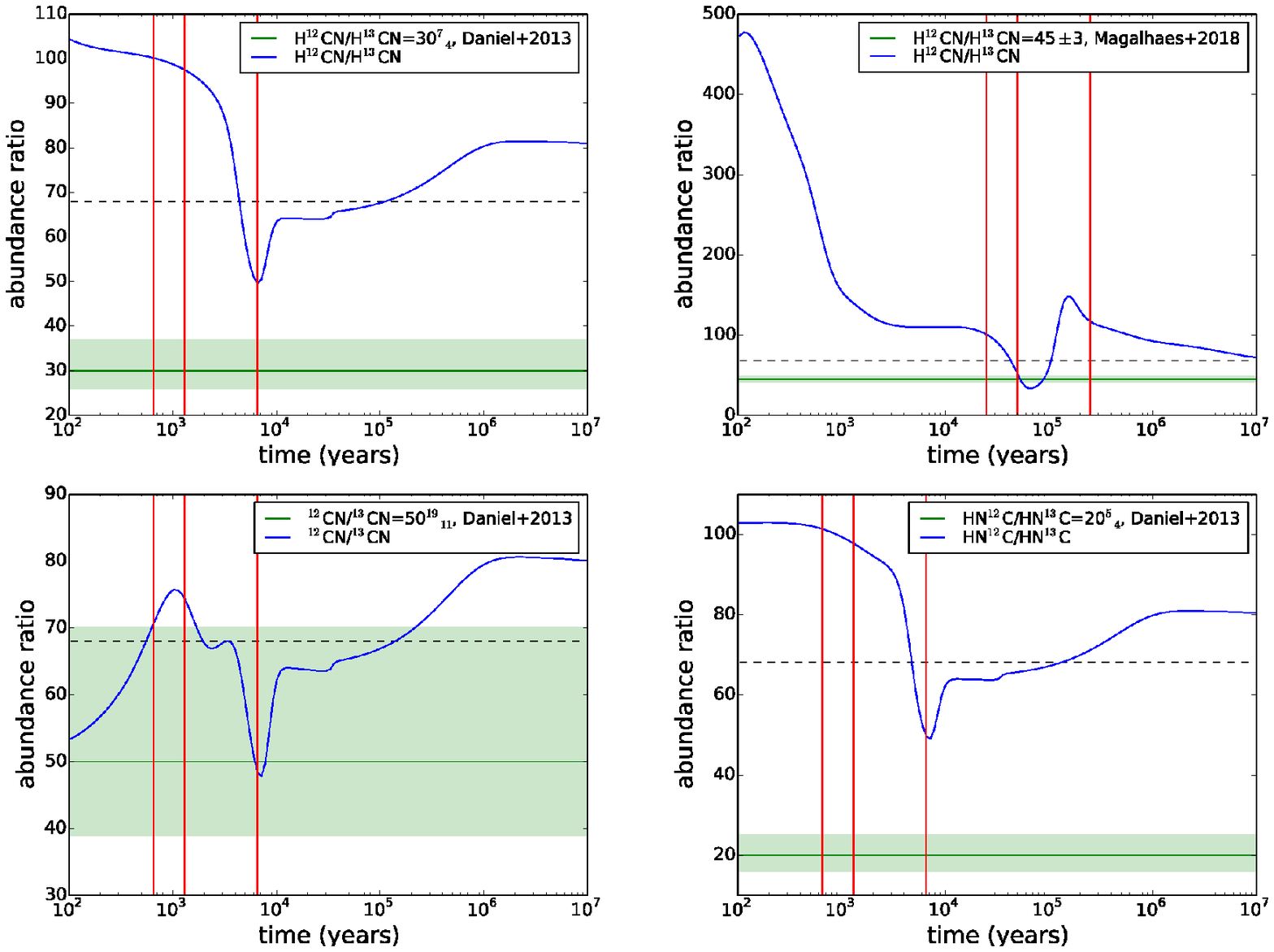}
\caption{\emph{Top right panel:} time evolution of the \cratio\;ratio for HCN for the model that simulates the center of the pre-stellar core L1498 observed by \citet{magalhaes2018} ($n_{\rm H}$=2$\times$10$^{5}$ cm$^{-3}$ and $T_{\rm gas}$=10 K). \emph{Top left panel and bottom panels:} time evolution of the \cratio\;ratio for HCN, CN, and HNC for the model that simulates the center of the pre-stellar core B1b observed by \citet{daniel2013} ($n_{\rm H}$=6$\times$10$^{6}$ cm$^{-3}$ and $T_{\rm gas}$=17 K). In all panels, the green horizontal line indicates the observed ratio, with the associated uncertainty as green area. The three red vertical lines represent $t_{\rm 1}$, $t_{\rm 2}$, and $t_{\rm 3}$. The black horizontal dashed line represents the initial \cratio\;ratio of 68.}
\label{fig-dan-mag-timeev}
\end{figure*}

As already discussed in the introduction of this chapter, \citet{daniel2013} found towards B1b HNC/HN$^{13}$C=20$^{\raisebox{1pt}{\footnotesize\rlap{\eqmakebox[subp][l]{+5}}}}_{\footnotesize\raisebox{1pt}{\eqmakebox[subp][l]{-4}}}$, HCN/H$^{13}$CN=30$^{\raisebox{1pt}{\footnotesize\rlap{\eqmakebox[subp][l]{+7}}}}_{\footnotesize\raisebox{1pt}{\eqmakebox[subp][l]{-4}}}$, and CN/$^{13}$CN=50$^{\raisebox{1pt}{\footnotesize\rlap{\eqmakebox[subp][l]{+19}}}}_{\footnotesize\raisebox{1pt}{\eqmakebox[subp][l]{-11}}}$, and \citet{magalhaes2018} towards the pre-stellar core L1498 obtained a HCN/H$^{13}$CN ratio of 45$\pm$3. \citet{daniel2013} derived a hydrogen density of 6$\times$10$^{6}$ cm$^{-3}$ and a temperature of 17 K towards the center of the pre-stellar core. Furthermore, \citet{magalhaes2018} found for the center of the core an average hydrogen density of 2$\times$10$^{5}$ cm$^{-3}$ and a temperature of 10 K. Thus, we use these physical properties to study the \cratio\;ratios predicted by our chemical model. The comparison is shown in Figs.~\ref{fig-HNC-parspaceexplor}, \ref{fig-HCN-parspaceexplor}, and \ref{fig-CN-parspaceexplor}.

For CN, we reproduce the values observed by \citet{daniel2013} at time $t_{\rm 3}$, while for HCN and HNC the values we found with our model are slightly higher. Note that the results of \citet{daniel2013} for HCN and HNC have large error bars because the HCN and HNC lines are heavily saturated. Opacity effects strongly impact the column density derivations and may lead to an underestimate of the main isotope column density and thus of the \cratio\;ratios. This could explain why we are not able to reproduce the values derived by \citet{daniel2013}.
Moreover, we can reproduce the value observed for HCN by \citet{magalhaes2018} at time $t_{\rm 2}$. This behaviour is also shown in Fig.~\ref{fig-dan-mag-timeev}, which represents the time evolution of the \cratio\;ratio for the two models corresponding to the two observed low-mass star-forming regions. Note that since the model is not simulating dynamical evolution, the fact that different observations could be reproduced at different times cannot be used to conclude anything about the chemical age of the simulated star-forming region. In addition, we recall that the initial conditions are taken somewhat arbitrarily so that the displayed time dependence is only indicative.

\subsubsection{High-mass star-forming regions sample}
\label{high-mass}

We used our grid of models to estimate the \cratio\;ratio for HNC and HCN towards the high-mass star-forming regions sample observed by \citet{colzi18a}. In particular, we used the H$_{2}$ column densities listed in Table~1 by \citet{fontani2018b} to derive the H$_{2}$ densities towards the region described by the SCUBA or APEX effective beam\footnote{The effective beam of the continuum observations is 22\asec\;for SCUBA and 15\asec\;for APEX. The instrument used for each source is given in Table~1 of \citet{fontani2018b}.}. The range of average $n_{\rm H}$ of the sources is from 2$\times$10$^{4}$ cm$^{-3}$ up to 6$\times$10$^{5}$ cm$^{-3}$. The kinetic temperatures, given in Table~3 of \citet{colzi18a}, are in between 14 and 47 K, and are similar to the range of values used in the grid of models performed earlier in this section.
With this information we can constrain the $^{12}$C/$^{13}$C ratios predicted by our models that could be used to compute N-fractionation from $^{13}$C-isotopologue observations. Figs.~\ref{fig-HNC-parspaceexplor} and \ref{fig-HCN-parspaceexplor} show the HN$^{12}$C/HN$^{13}$C and H$^{12}$CN/H$^{13}$CN ratios, for the three analysed times, together with the observed volume density range.

The \cratio\;ratios derived for $t_{\rm 1}$, $t_{\rm 2}$, and $t_{\rm 3}$, and the errors made when the \nratio\;ratios are measured assuming a fixed local \cratio\;ratio of 68, for HNC and HCN, are listed in Table \ref{table-highmassratio-factors}.
Taking, for example, one of the sources with the highest H$_{2}$ number density, 19410+2336, with $n_{\rm H}$=5.8$\times$10$^{5}$ cm$^{-3}$ and $T_{\rm gas}$=21 K, \citet{colzi18a} derived a HNC/H$^{15}$NC ratio of 431$\pm$24. For the early chemistry time we obtain HNC/H$^{15}$NC$\sim$550. For $t_{\rm 2}$ it is $\sim$495, and for $t_{\rm 3}$ the ratio goes down to $\sim$370. This shows that our results are time-dependent, thus care needs to be taken in the comparison with observations, because the chemical times may not correspond to the dynamical age of the star-forming region.
So, a more accurate model including also the evolution of physical parameters is necessary to better constrain the time-scales. Moreover, the H$_{2}$ column densities are derived from regions (15\asec\;or 22\asec) that probably also include the more diffuse gas that surrounds the denser cores in which star-formation occurs. As an example, \citet{beuther2007b} derived an H$_{2}$ column density of $\sim$10$^{23}$--10$^{24}$ cm$^{-2}$ in regions of $\sim$2\asec\;in size towards the denser cores of the high-mass protocluster IRAS 05358 (see also \citealt{colzi2019}). These column densities correspond to H$_{2}$ number densities of $\sim$10$^{6}$--10$^{7}$ cm$^{-3 }$ that are one or two orders of magnitude higher than those used in this section to compare with the model. Thus, detailed continuum observations and a description of the structures of these regions are needed.

\begin{table}
\normalsize
\setlength{\tabcolsep}{8pt}
\caption{\cratio\;ratios predicted for HNC and HCN in the range of temperatures and densities of the high-mass star-forming regions sample observed by \citet{colzi18a} (Cols. 2 and 3). The fourth and fifth columns present the error factor made when deriving \nratio\;ratios assuming a local \cratio\;ratio of 68.}
  \begin{tabular}{lcccc}
  \hline
  Time   & $R_{1}=\frac{{\rm HN}^{12}{\rm C}}{{\rm HN}^{13}{\rm C}}$ & $R_{2}=\frac{{\rm H}^{12}{\rm CN}}{{\rm H}^{13}{\rm CN}}$ & $R_{1}$/68 &$R_{2}$/68\\ 
\hline
 $t_{1}$  & 74--135 &68--121 &1.1--1.9 & 1--1.8 \\
  $t_{2}$     & 68--97 & 52--96 & 1--1.4 & 0.8--1.4 \\
  $t_{3}$      &53--240  & 52--210 & 0.8--3.5& 0.8--3.1 \\   
  \hline
  \normalsize
  \label{table-highmassratio-factors}
  \end{tabular}
  \centering
\end{table}

\subsubsection{Effect of cosmic rays}
\label{effectCR}

\begin{figure*}
\centering
\includegraphics[width=45pc]{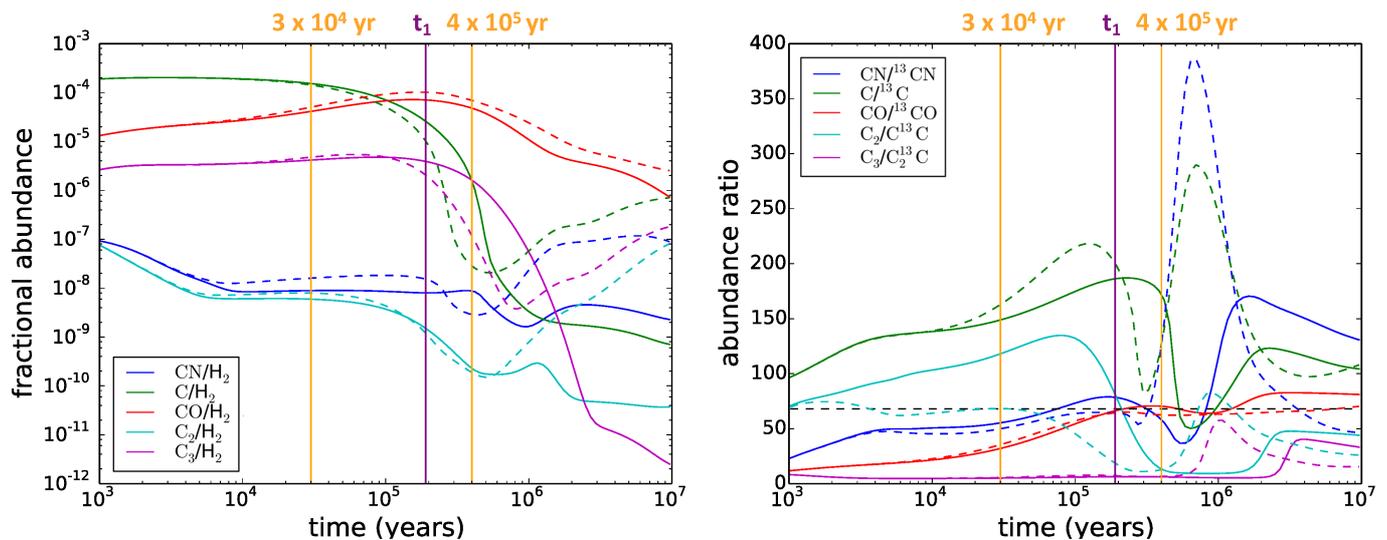}
\caption{\emph{Left panel:} Time evolution of CN, C, CO, C$_{2}$, and C$_{3}$ abundances with respect to H$_{2}$ for the fiducial model with a cosmic-ray ionization rate of 1.3$\times$10$^{-18}$ s$^{-1}$. \emph{Right panel:} Time evolution of the \cratio\;ratio for CN, C, CO, C$_{2}$, and C$_{3}$ for the fiducial model with a cosmic-ray ionization rate of 1.3$\times$10$^{-18}$ s$^{-1}$. In both panels, the vertical purple solid line represents the "early chemistry" time as defined in the text, while the two vertical orange solid lines represent the two times (3$\times$10$^{4}$ yr, and 4$\times$10$^{5}$ yr) that we have analysed and discussed in the text. The trends obtained with the standard $\zeta$ of the fiducial model are superimposed with dashed lines. The black horizontal dashed line represent the initial \cratio\;ratio of 68.}
\label{fig-C3-behaviou-lowerz}
\end{figure*}

\begin{figure*}
\centering
\includegraphics[width=45pc]{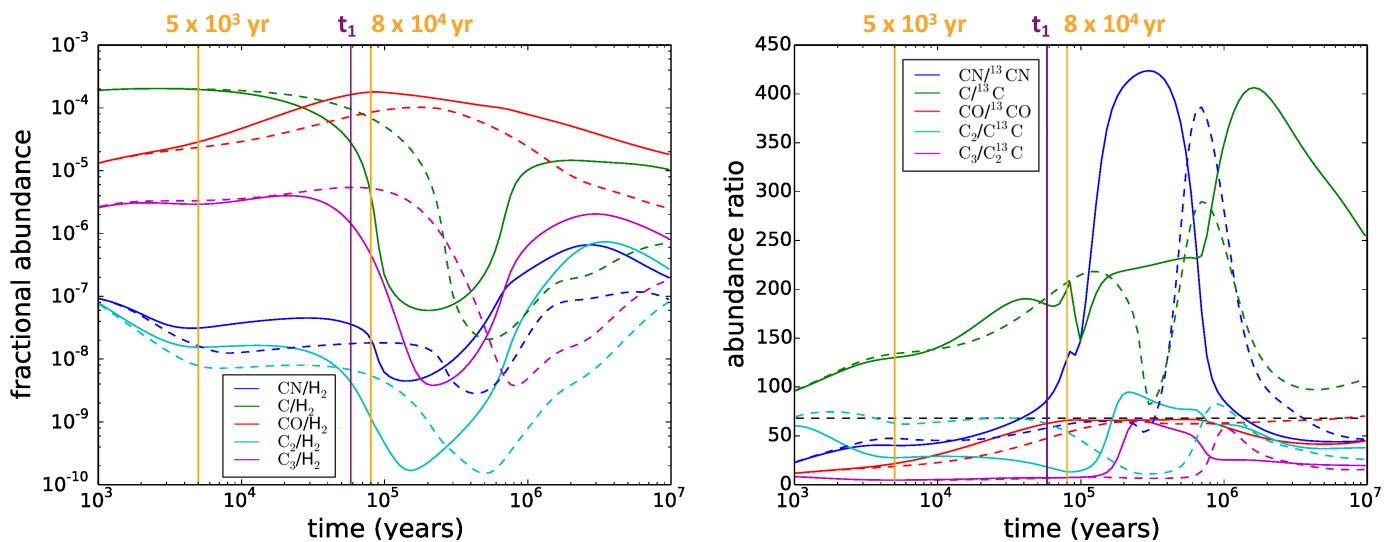}
\caption{\emph{Left panel:} As Fig.~\ref{fig-C3-behaviou-lowerz}, but for $\zeta$ = 1.3$\times$10$^{-16}$ s$^{-1}$. In both panels, the vertical purple solid line represents the "early chemistry" time as defined in the text, while the two vertical orange solid lines represent the two times (5$\times$10$^{3}$ yr, and 8$\times$10$^{4}$ yr) that we have analysed and discussed in the text. The trends obtained with the standard $\zeta$ of the fiducial model are superimposed with dashed lines. The black horizontal dashed line represent the initial \cratio\;ratio of 68.}
\label{fig-C3-behaviour-higherz}
\end{figure*}

\begin{figure*}
\centering
\includegraphics[width=36pc]{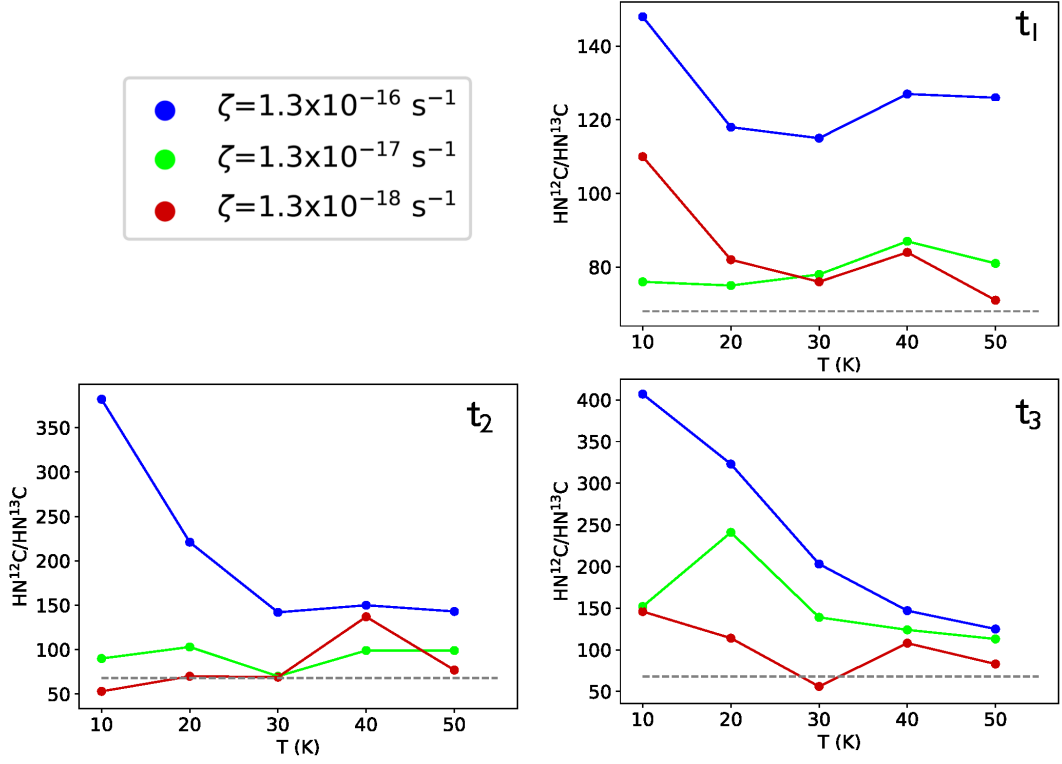}
\caption{HN$^{12}$C/HN$^{13}$C ratio as a function of the temperature, for different cosmic-ray ionization rates, at $t_{\rm 1}$ (\emph{top right panel}), $t_{\rm 2}$ (\emph{bottom left panel}), and $t_{\rm 3}$ (\emph{bottom right panel}). The black horizontal dashed line represent the initial \cratio\;ratio of 68.}
\label{fig-HNC-cosmicrays}
\end{figure*}

\begin{figure*}
\centering
\includegraphics[width=36pc]{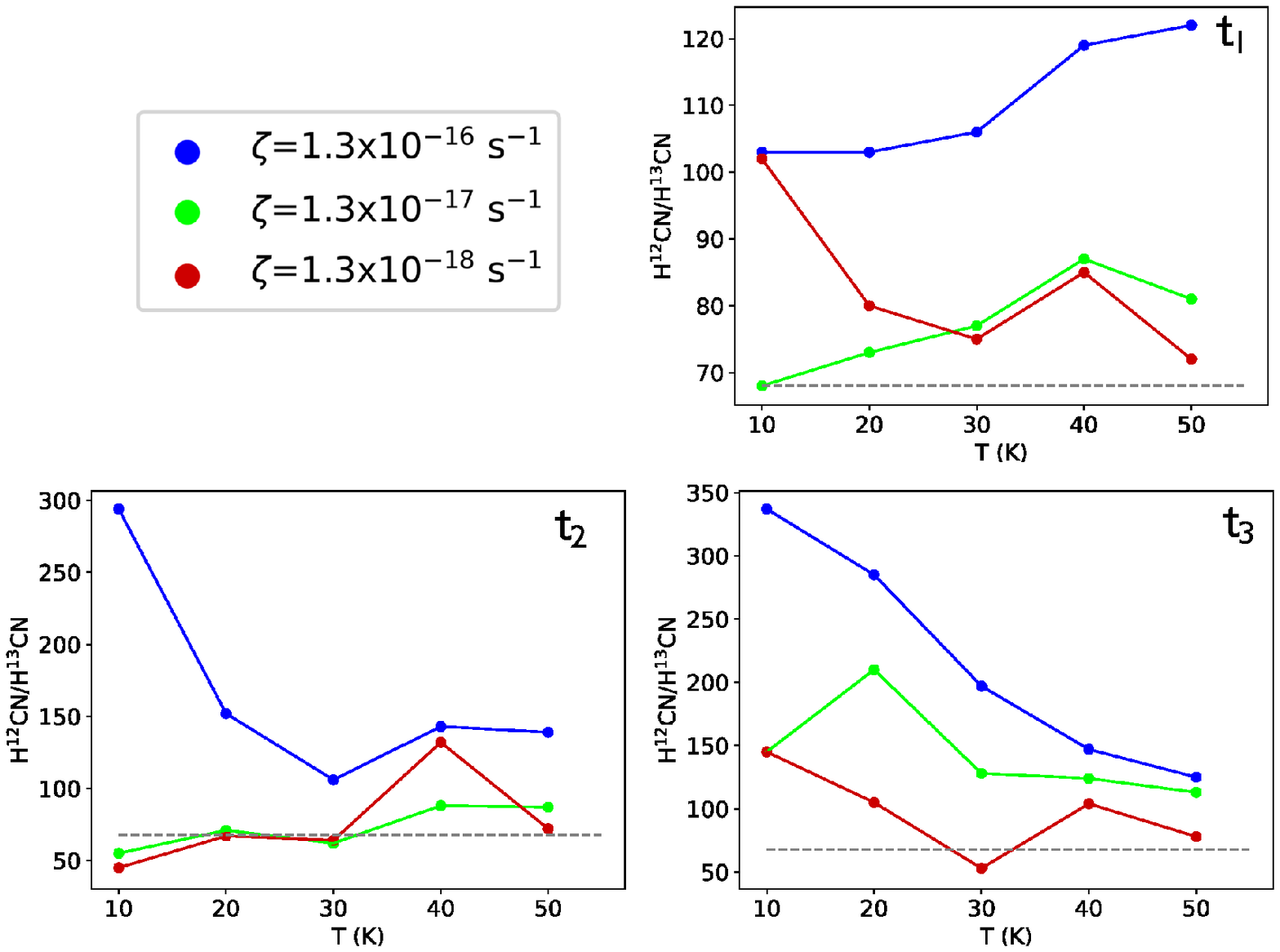}
\caption{As Fig.~\ref{fig-HNC-cosmicrays}, but for H$^{12}$CN/H$^{13}$CN.}
\label{fig-HCN-cosmicrays}
\end{figure*}

\begin{figure*}
\centering
\includegraphics[width=36pc]{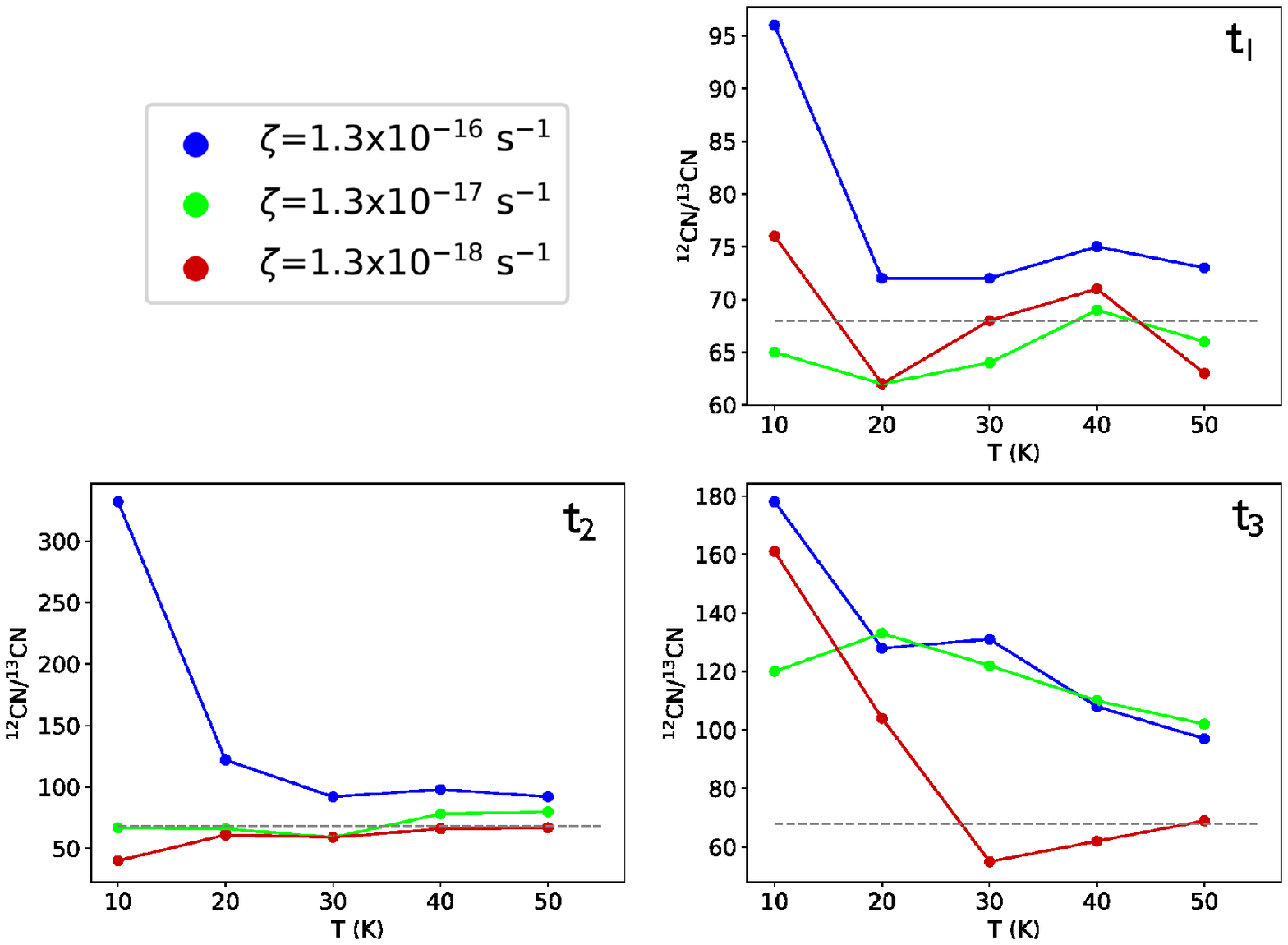}
\caption{As Fig.~\ref{fig-HNC-cosmicrays}, but for $^{12}$CN/$^{13}$CN.}
\label{fig-CN-cosmicrays}
\end{figure*}

Finally, we discuss the possible effect of different cosmic-ray ionization rates ($\zeta$) in the fiducial model.
Until now we used the canonical value of 1.3$\times$10$^{-17}$ s$^{-1}$ (e.g. \citealt{padovani2009}). We analyse here how the main formation and destruction routes (cf. Fig.~\ref{fig-13C-reactions}) change when one assumes lower and higher $\zeta$ with respect to the standard one: $\zeta_{\rm low}$=1.3$\times$10$^{-18}$ s$^{-1}$ and $\zeta_{\rm high}$=1.3$\times$10$^{-16}$ s$^{-1}$, respectively. For example, $\zeta_{\rm high}$ is similar to the value derived in the diffuse cloud in the vicinity of OMC-2 FIR4 by \citet{lopez-sepulcre2013} (2.3$\times \zeta_{\rm high}$), and \citet{caselli1998} used deuterium fraction measurements of HCO$^{+}$ to constrain $\zeta$ in the range of 10$^{-18}$ to 10$^{-16}$~s$^{-1}$ towards 24 cloud cores.

 In the case of $\zeta_{\rm low}$, the early chemistry time is 1.9$\times$10$^{5}$ yr, and in the model with $\zeta_{\rm high}$ it is reduced to 5.8$\times$10$^{4}$ yr.
Figs.~\ref{fig-C3-behaviou-lowerz} and \ref{fig-C3-behaviour-higherz} show the behaviour of the abundances and \cratio\;ratios of the same molecules studied for the fiducial model, for $\zeta_{\rm low}$ and $\zeta_{\rm high}$, respectively. For these two models, we studied in detail two time-scales, before and after the early chemistry time. Figs.~\ref{fig-13C-reactions-lowerz} and \ref{fig-13C-reactions-higherz} represent the main destruction and formation pathways of various species for $\zeta_{\rm low}$ and $\zeta_{\rm high}$, respectively. 

In the case of $\zeta_{\rm low}$:

\begin{itemize}

\item For $t<t_{\rm 1}$, $^{12}$C is very abundant, as in the standard case, because of the efficiency of the low-temperature isotopic exchange reactions. However, in this model the C$_{2}$/C$^{13}$C ratio is also higher than 68 and similar to the C-fractionation of atomic carbon. In fact, the secondary photon reactions are less efficient in creating C$^{13}$C starting from C$_{2}^{13}$C.\\

\item For $t>t_{\rm 1}$, the isotopic exchange reaction involving C$_{3}$ is still efficient since the destruction reaction of C$_{2}^{13}$C with secondary photons is not as important as in the fiducial model. However, the atomic \cratio\;ratio decreases with the depletion of $^{12}$C onto grain surfaces.
Moreover, the abundances of atomic carbon and related molecules (like C$_{3}$ and CN) do not increase again in the late chemistry since He$^{+}$ is not efficiently produced by cosmic-ray reactions owing to the low $\zeta$. This also affects the decrease of the \cratio\;ratios.
\end{itemize}

In the case of $\zeta_{\rm high}$:
\begin{itemize}
\item For $t<t_{\rm 1}$, reactions with secondary photons are the most important ones. Thus, both $^{12}$C- and $^{13}$C-containing species follow the same chemical pathway (see left panel of Fig.~\ref{fig-13C-reactions-higherz}). In this case, C$_{2}$ inherits the \cratio\;ratio from C$_{3}$ (which is lower than 68), and the C-fractionation of atomic carbon and C$_{3}$ is still governed by the very efficient low-temperature isotopic exchange reaction.
\item For $t>t_{\rm 1}$, the behaviour is similar to the fiducial model. The main difference is that the isotopic exchange reaction of C$_{3}$ is still efficient since secondary photons create more atomic $^{13}$C (from C$_{2}$ whose C-fractionation is inherited by C$_{3}$) than in the fiducial model, maintaining the efficiency of this reaction for a longer time. As a consequence, the atomic \cratio\;ratio remains higher and there is no decreasing trend toward values lower than 68, not even for the molecules related to atomic carbon (CN, HCN, HNC).
In the late chemistry time, atomic carbon and C-chains form very efficiently because of the enhanced abundance of He$^{+}$ due to the higher cosmic-ray ionization rate. This effect maintains a higher abundance of atomic carbon, and consequently the atomic \cratio\;ratio is larger.
\end{itemize}

We also performed a parameter-space exploration for a fixed $n_{\rm H}$ of 2$\times$10$^{4}$ cm$^{-3}$, varying the temperatures from 10 up to 50~K and the cosmic-ray ionization rate between $\zeta_{\rm low}$ and $\zeta_{\rm high}$. In particular, as already done in Sect.~\ref{parspace}, we focused the analysis on HNC, HCN, and CN (Figs.~\ref{fig-HNC-cosmicrays}, \ref{fig-HCN-cosmicrays}, and \ref{fig-CN-cosmicrays}). The main trend is that the \cratio\;ratios are highest for $\zeta_{\rm high}$. This is related to the fact that secondary photon reactions maintain a high efficiency of low-temperature isotopic exchange reactions that involve C$_{3}$, and the atomic \cratio\;ratio (and that of related molecules) remains higher than 68 (see also Fig.~\ref{fig-13C-reactions-higherz}, right panel).

\section{Conclusions}

We developed a new chemical network to study in detail how important isotopic exchange reactions are to the chemistry of carbon-containing species in the low temperature (10--50 K) gas in star-forming regions with $n_{\rm H}$=2$\times$10$^{3}$--2$\times$10$^{7}$ cm$^{-3}$. 
In particular, we suggest the occurrence of $^{13}$C exchange involving C$_{3}$ with atomic carbon and study the possible consequences of this reaction.
The main results and conclusions of this work are summarised below:
\begin{enumerate}
\item We found that reaction \eqref{reac-C3} is mainly important for $T_{\rm gas}<$27~K (owing to its exothermicity), and leads to \cratio$<$68 for molecules that form from atomic carbon (e.g. C$_{2}$, C$_{3}$, and nitrile-bearing species). This behaviour occurs in the time period between the conversion of atomic carbon into CO, and the time when CO is almost completely frozen-out onto dust grains. \\
\item We performed a detailed study of the \cratio\;ratios for the nitrile-bearing species HCN, HNC, and CN, as a function of density and for different temperatures. Our model can partially reproduce the carbon isotopic ratios derived by \citet{daniel2013} and \citet{magalhaes2018} from observations towards low-mass pre-stellar cores.\\
\item We also used our grid of models to evaluate how far we are from the correct \cratio\;ratio, assumed to be 68 in the local ISM, to derive the \nratio\;ratios in the sample of high-mass star-forming regions analysed by \citet{colzi18a}. Our chemical model predicts that we are over- or underestimating the assumed \cratio\;ratios by factors $\sim$0.8--1.9, even up to $\sim$3.5, depending on the evolutionary time that we choose for the analysis. Considering these factors, the average nitrogen fractionation value of $\sim$330 derived by \citet{colzi18a} and \citet{colzi18b} for HCN and HNC, could become 260--1150. This range would include both the low values measured in comets (e.g. 250), and the higher values found in low-mass pre-stellar cores (e.g. $\sim$630--770 for N$_{2}$H$^{+}$ by \citealt{redaelli2018}). This highlights the importance of knowing the precise \cratio\;ratios in studies of nitrogen fractionation with the double-isotope method.\\
\item Finally, we performed a parameter-space exploration of the \cratio\;ratios for nitrile-bearing species, varying the cosmic-ray ionization rate ($\zeta$). We found that for $\zeta$ higher than the standard value of 1.3$\times$10$^{-17}$ s$^{-1}$, the \cratio\;ratios are on average higher than in the cases with lower $\zeta$. This result is due to the importance of secondary photon reactions in maintaining efficient the isotopic exchange reactions involving C$_{3}$, and to cosmic rays, which lead to an efficient production of atomic carbon and C-chains.
\end{enumerate}

In a future work, we will upgrade the chemical network to cover nitrogen fractionation as well. In this way we will be able to better describe the interplay between carbon and nitrogen isotope chemistry. The inclusion of the position of $^{13}$C in multiple carbon bearing species is also another challenge left for future work.

\begin{acknowledgements}
Many thanks to the anonymous referee for the careful reading of the paper and the comments that improved the work.
This publication has received funding from the European Union’s Horizon 2020 research and innovation programme under grant agreement No 730562 [RadioNet].
LC acknowledges support from the Italian Ministero dell’Istruzione, Università e Ricerca through the grant Progetti Premiali 2012 - iALMA (CUP C52I13000140001). PC acknowledges support from the European Research Council (project PALs 320620). ER acknowledges partial support by the Programme National “Physique et Chimie du Milieu Interstellaire” (PCMI) of CNRS/INSU with INC/INP co-funded by CEA and CNES. Most of the figures were generated with the PYTHON-based package MATPLOTLIB (\citealt{hunter2007}).
\end{acknowledgements}

\bibliographystyle{aa} 
 \bibliography{bibliography} 

 \begin{appendix}
 \clearpage
\section{Comparison with existing fractionation procedure and chemical models}

In this Appendix, we compare the results of our chemical model with other results already published in the literature. Firstly, in Sect.~\ref{compar-roueff}, we analyse the fractionation procedures in the implementation of the $^{13}$C chemistry and compare with the method used in \citet{roueff2015}. Secondly, in Sect.~\ref{compar-furu} we compare our results to the \cratio\;ratios predicted by \citet{furuya2011}.

\subsection{Comparison with \citet{roueff2015} fractionation procedure}
\label{compar-roueff}

The settling of fractionation reactions involving specific isotopic exchange represents a first important step which has to be completed by the fractionation of the various reactions involving carbon containing molecules, as described in Sect.~\ref{frac-procedure}.
Such a procedure is not unequivocal when several carbon atoms are involved in a chemical reaction. We have carefully analysed and compared the present method to the study performed by \citet{roueff2015} for three different initial conditions, in order to test the robustness of our findings. In particular, we have used the initial conditions of this work (fiducial model with $n_{\rm H}$=2$\times$10$^{4}$~cm$^{-3}$ and a temperature of 10 K), and those for model (a) and model (b) described in Table 4 of \citet{roueff2015} (starless and pre-stellar core conditions, respectively). The results of the comparison are shown in Tables \ref{table-comparison-roueff-sipila}, \ref{table-comparison-roueff-modela}, and \ref{table-comparison-roueff-modelb}, for the initial conditions of this work, model (a) and model (b), respectively.
The tests are performed on the steady state results of the gas-phase model, as \citet{roueff2015} did not consider surface reactions. Moreover, they restricted carbon-containing molecules to three carbon atoms, and included up to one $^{13}$C in each molecule.

We first ran the chemical network used in \citet{roueff2015} by suppressing the substituted $^{15}$N nitrogen species but keeping the deuterated components. 
We realized that reactions involving two carbon containing molecules, which could lead to two other different carbon containing molecules, could be fractionated differently, according to the chemical mechanism involved.
Let us consider the reaction AC + BC $\rightarrow$ A'C + B'C.
The reaction A$^{13}$C + BC may then lead to A'$^{13}$C + B'C if the reaction proceeds directly, with the same reaction rate coefficient than for the initial reaction. However, if an intermediate complex is occurring, a redistribution of $^{13}$C can take place and the reaction may lead to a complete scrambling of the carbon atoms. In this case an additional channel A'C + B'$^{13}$C can be open and then we assume that each channel occurs with the half value of the rate. The fractionation procedure used in \citet{roueff2015} did not include this possibility. We thus compare in columns 1 and 2 of the Tables in this section the role of such scrambling possibilities in the \cratio\;ratios of several observed molecules. For example, Table \ref{table-comparison-roueff-sipila} shows that some ratios are more sensitive than others to this assumption, H$_{2}$CO and HNC being particularly sensitive. The introduction of full scrambling leads to a reduction of the \cratio\;ratio when the $^{13}$C isotopologue is found to be diluted in the non-scrambling approach. However, the lower ratios (e.g. for CO, HCO$^{+}$) are not affected by this mechanism.

An additional test was performed by using the same chemical network with the additional assumption that the fractionation of $^{13}$C acts only on molecules containing a single carbon. The third column reports the results from this model, including systematic scrambling as described previously. The fourth column displays the results obtained with the chemical network used in this work, described in Sect.~\ref{frac-procedure}, assuming as above that the fractionation of $^{13}$C acts only on molecules containing a single carbon. We see that the results are in a reasonable agreement with some small differences. These can be attributed to the additional sampling of the reactions made in the present study where charge exchange reactions are quoted specifically. An additional difference between the two treatments concerns the omission of the reactions between two $^{13}$C substituted reactants in the test results reported in column 4, whereas these reactions are taken into account in the tests reported in columns 1, 2 and 3.
All together, we conclude that the main trends of the isotopic ratios are preserved in the different hypotheses made to build the isotopic chemistry but we indeed acknowledge the unavoidable remaining uncertainties of any systematic procedure.

\begin{table*}
\normalsize
\caption{Comparison between the fractionation procedure used in this work and the one by \citet{roueff2015}: steady state \cratio\;ratios of different molecular species in a chemical model adopting the initial conditions of the fiducial model of this work.}
  \begin{tabular}{lcccccc}
  \hline
  \cratio\;ratio   & without scrambling & full scrambling & 1C-molecule & 1C-molecule (this work) \\ 
  \hline
\cratio  & 69.0              & 69.0            & 69.7           &  75.8   \\
 $^{12}$CH/$^{13}$CH & 296.2         & 277.0           & 287.0       & 304.0\\
  $^{12}$CO/$^{13}$CO & 68.2           &68.2            &68.2           &68.4 \\
   H$_{2}^{12}$CO/H$_{2}^{13}$CO &  304.6         & 195.6        &299.6          & 307.5 \\
 $^{12}$CN/$^{13}$CN & 104.3          &90.6           & 204.2         & 287.4 \\
  H$^{12}$CN/H$^{13}$CN & 271.0      & 130.6        & 299.3        & 283.5\\
 HN$^{12}$C/HN$^{13}$C& 295.1        & 128.6         & 321.6          &283.8 \\
  $^{12}$CS/$^{13}$CS & 85.5           &78.1            &87.0          & 244.6\\
   H$^{12}$CO$^{+}$/H$^{13}$CO$^{+}$ &28.3          &28.3            & 28.3           & 26.0 \\
   \hline
  \normalsize
  \label{table-comparison-roueff-sipila}
  \end{tabular}
\centering
\tablefoot{Column 1 represents the results obtained with the same chemical network of \citet{roueff2015} by suppressing the substituted $^{15}$N nitrogen species but keeping the deuterated components, and without considering a full scrambling of carbon atom in molecules, as described in Sect.~\ref{compar-roueff}. Column 2 is as Column 1 but considering full scrambling. Column 3 represents the results obtained with the same network as Column 2, but considering inclusion of $^{13}$C only in molecules containing one carbon atom. Column 4 shows the corresponding results obtained with the chemical network developed for this work (Sect.~\ref{frac-procedure}), reduced with the same assumption made for Column 3.}
      \normalsize
\end{table*}

\begin{table*}
\normalsize
\caption{As Table \ref{table-comparison-roueff-sipila}, but adopting the initial conditions of model (a) in \citet{roueff2015}, corresponding to a starless core.}
  \begin{tabular}{lcccccc}
  \hline
  \cratio\;ratio   & without scrambling & full scrambling &  1C-molecule & 1C-molecule (this work)\\ 
  \hline
\cratio  & 102.1             & 101.7           &115.7           & 193.5  \\
 $^{12}$CH/$^{13}$CH & 487.2           &465.3          & 412.1         & 416.9 \\
  $^{12}$CO/$^{13}$CO & 68.0          & 68.0          & 67.8           & 68.0 \\
   H$_{2}^{12}$CO/H$_{2}^{13}$CO &  490.5           & 382.9         & 422.6           &  418.1\\
 $^{12}$CN/$^{13}$CN & 216.9         & 191.5         & 291.8         & 402.4  \\
  H$^{12}$CN/H$^{13}$CN & 266.7         & 198.4         & 284.0        & 276.5\\
 HN$^{12}$C/HN$^{13}$C& 345.6         & 212.8         & 321.3        &  273.8  \\
  $^{12}$CS/$^{13}$CS & 144.0         &135.3       & 159.1        &  364.2\\
   H$^{12}$CO$^{+}$/H$^{13}$CO$^{+}$ &45.5        &45.6      & 47.1         & 40.6 \\
   \hline
  \normalsize
  \label{table-comparison-roueff-modela}
  \end{tabular}
\centering
      \normalsize
\end{table*}

\begin{table*}
\normalsize
\caption{As Table \ref{table-comparison-roueff-sipila}, but adopting the initial conditions of model (b) in \citet{roueff2015}, corresponding to a pre-stellar core.}
  \begin{tabular}{lcccccc}
  \hline
  \cratio\;ratio   & without scrambling & full scrambling & 1C-molecule & 1C-molecule (this work) \\ 
  \hline
\cratio  & 136.7          & 135.4         & 171.1        &  263.7   \\
 $^{12}$CH/$^{13}$CH & 445.9         & 432.3       & 380.2         & 343.6 \\
  $^{12}$CO/$^{13}$CO & 67.8           & 67.8          & 67.4          & 68.0 \\
   H$_{2}^{12}$CO/H$_{2}^{13}$CO &  443.3          &334.0          &382.4          & 342.5 \\
 $^{12}$CN/$^{13}$CN & 329.3          & 204.4         & 332.4        & 337.1 \\
  H$^{12}$CN/H$^{13}$CN & 292.5         &162.2       & 290.6         & 251.8\\
 HN$^{12}$C/HN$^{13}$C& 304.4         & 147.2        & 285.2       & 247.2  \\
  $^{12}$CS/$^{13}$CS & 230.0         & 171.8       & 257.1          &327.1\\
   H$^{12}$CO$^{+}$/H$^{13}$CO$^{+}$ &44.5           &44.5        & 43.7           & 39.4   \\
   \hline
  \normalsize
  \label{table-comparison-roueff-modelb}
  \end{tabular}
\centering
      \normalsize
\end{table*}

\subsection{Comparison with the previous model of \citet{furuya2011}}
\label{compar-furu}
\begin{figure*}
\centering
\includegraphics[width=45pc]{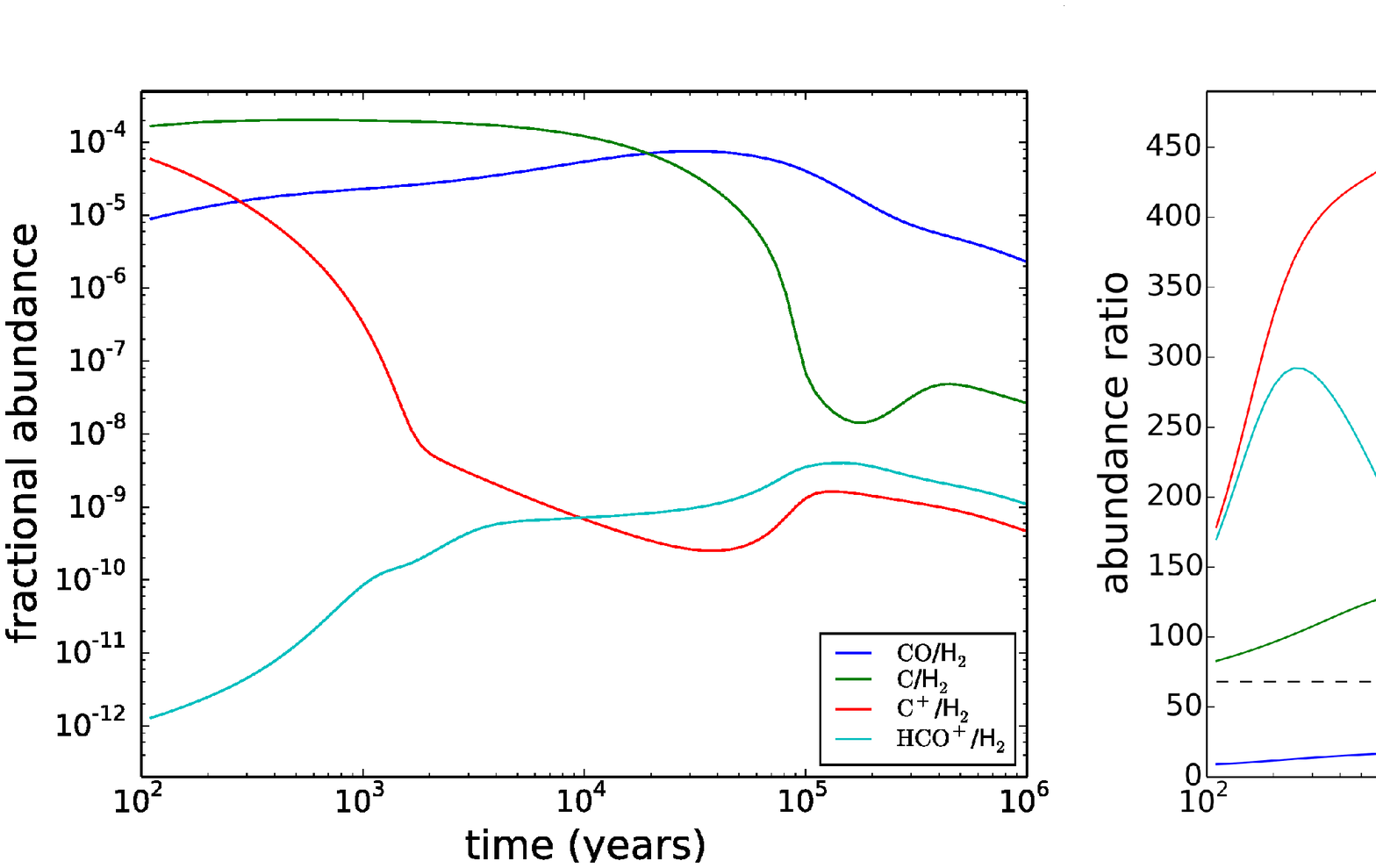}
\caption{Abundances (\emph{left panel}) and \cratio\;ratios (\emph{right panel}) from our chemical model using a density of 10$^{5}$ cm$^{-3}$ and temperature of 10 K (same as \citealt{furuya2011}). In the right panel the black horizontal dashed line represents the initial \cratio\;ratio of 68.}
\label{fig-comparison-furuya}
\end{figure*} 

\citet{furuya2011} studied the behaviour of \cratio\;for CCH and CCS with a gas-grain chemical model. In particular, their work was based on the study of isotopomer fractionation, that is the abundance ratio between the variants of a given species where the $^{13}$C position is allowed to vary. They introduced as isotopic exchange reactions only (1) and (2) of Table \ref{tab-reaccolzi}. Moreover, they did not consider multiple $^{13}$C species, contrary to our fractionation procedure that includes up to three $^{13}$C in molecules. Figure~\ref{fig-comparison-furuya} shows the abundances and \cratio\;ratios predicted by the present model for the same molecules studied by \citet{furuya2011} in their Fig. 1, for the same physical conditions. We did not include HC$_{3}$N since we cut the chemistry to 5 atoms-containing molecules and the results for this molecule could be biased by this assumption. The general time-dependence of CO, C, C$^{+}$, and HCO$^{+}$ is well reproduced in our model. The small differences are likely to be due to the introduction of new possible isotopic exchange reactions in our model that are efficient at low temperatures (like those for CN, C$_{2}$, CS, and C$_{3}$).

 \clearpage

\section{Parameter-space exploration of CO, HCO$^{+}$, H$_{2}$CO}
\label{parspace-other}
\begin{figure*}
\centering
\includegraphics[width=36pc]{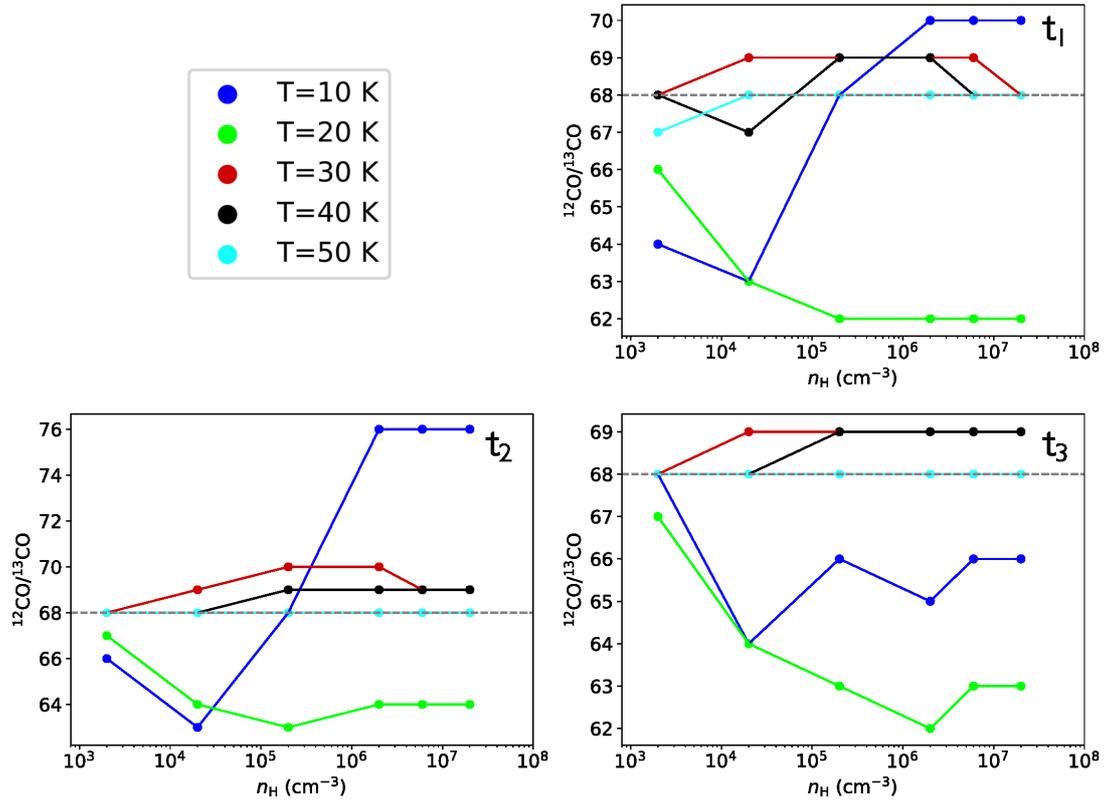}
\caption{$^{12}$CO/$^{13}$CO ratio as a function of $n_{\rm H}$, for different temperatures, at $t_{\rm 1}$ (\emph{top right panel}), $t_{\rm 2}$ (\emph{bottom left panel}), and $t_{\rm 3}$ (\emph{bottom right panel}). The black horizontal dashed line represents the initial \cratio\;ratio of 68.}
\label{fig-CO-parspaceexplor}
\end{figure*}

\begin{figure*}
\centering
\includegraphics[width=36pc]{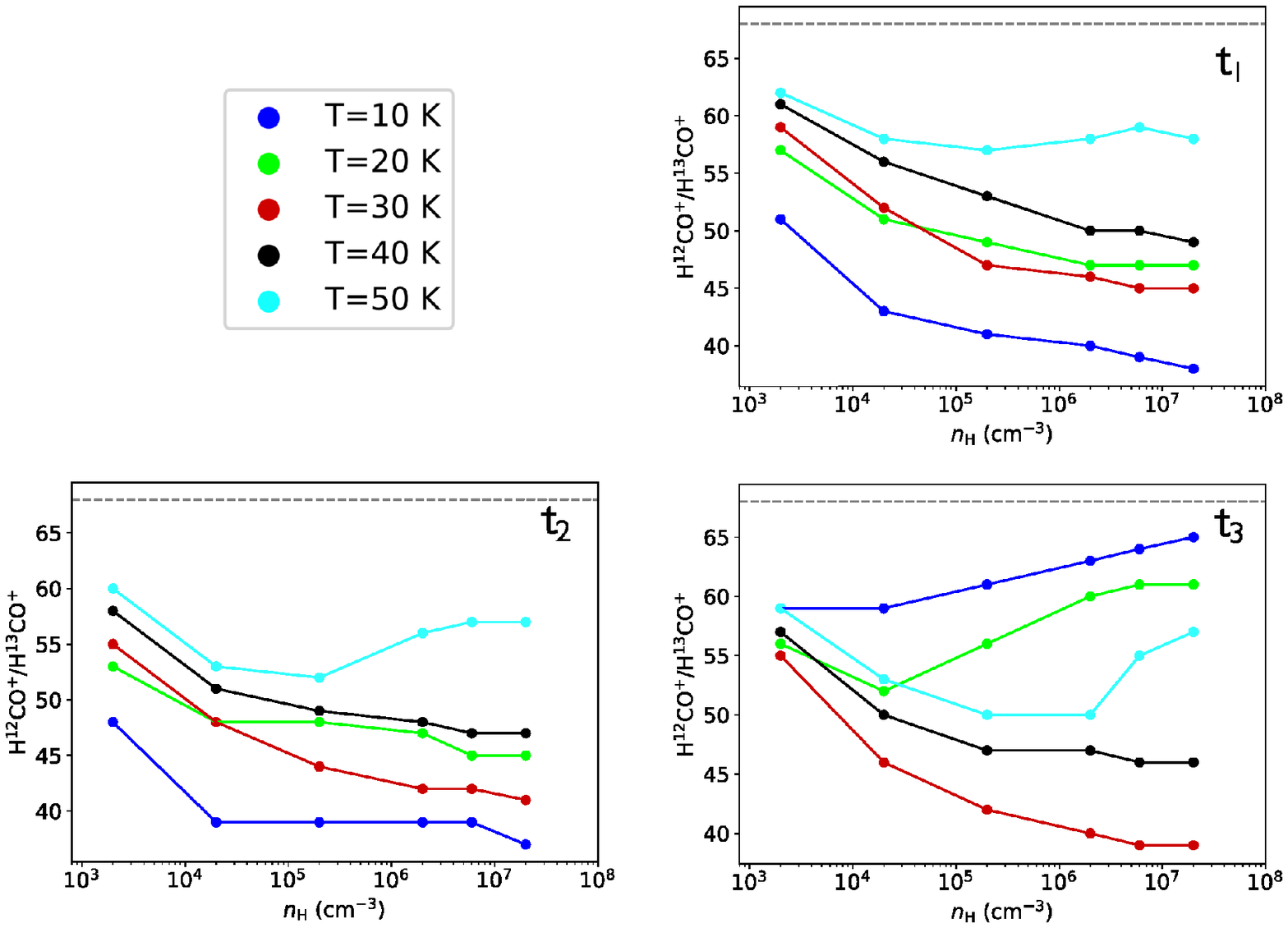}
\caption{As Fig.~\ref{fig-CO-parspaceexplor}, but for H$^{12}$CO$^{+}$/H$^{13}$CO$^{+}$.}
\label{fig-HCO+-parspaceexplor}
\end{figure*}

\begin{figure*}
\centering
\includegraphics[width=36pc]{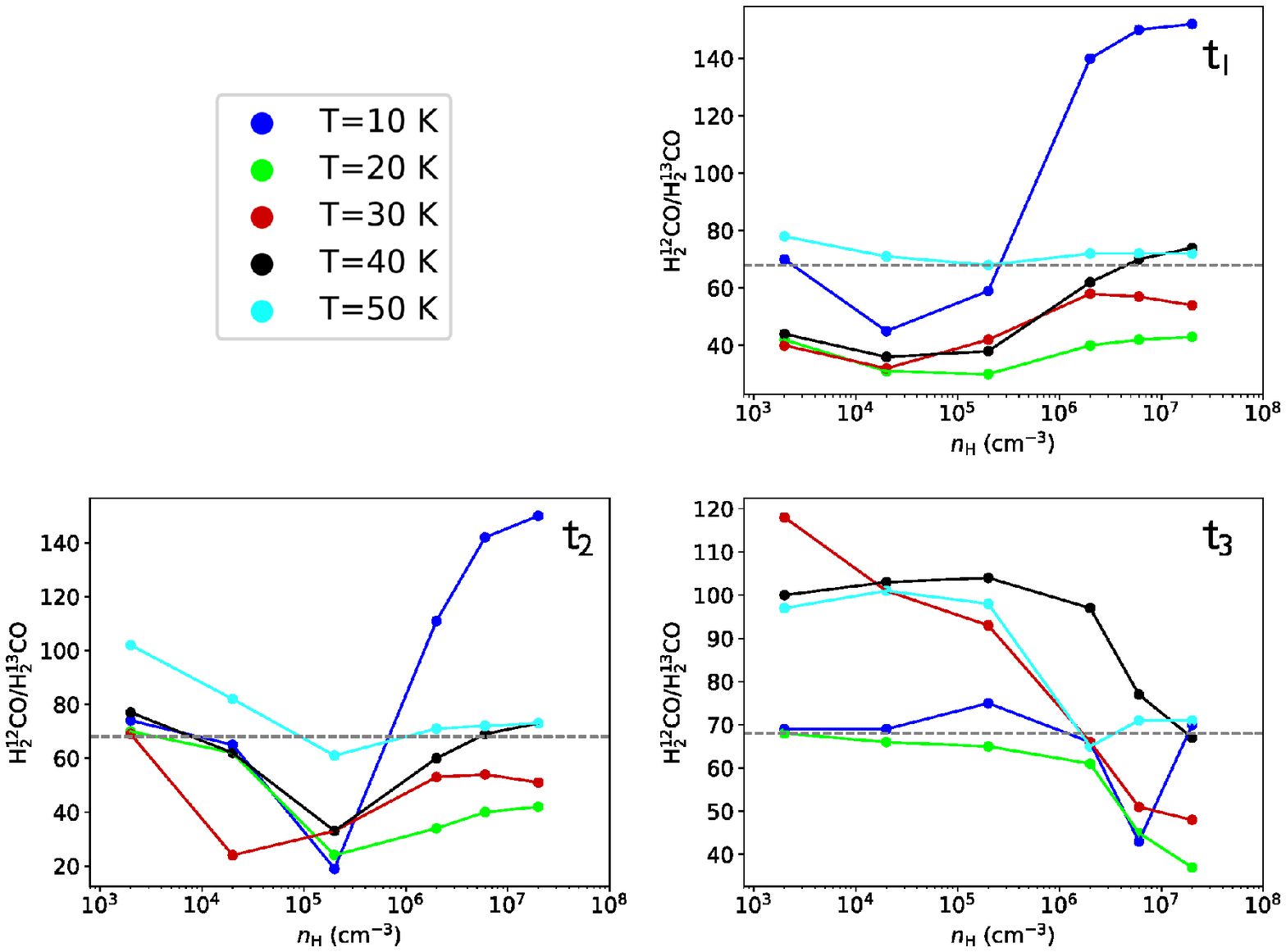}
\caption{As Fig.~\ref{fig-CO-parspaceexplor}, but for H$_{2}^{12}$CO/H$_{2}^{13}$CO.}
\label{fig-H2CO-parspaceexplor}
\end{figure*}

In this Appendix we show the \cratio\;ratio of CO, HCO$^{+}$, and H$_{2}$CO as a function of $n_{\rm H}$ and for different temperatures (Figs.~\ref{fig-CO-parspaceexplor}, \ref{fig-HCO+-parspaceexplor}, and \ref{fig-H2CO-parspaceexplor}, respectively). In particular, we perform the same parameter-space exploration as that done for CN, HCN, and HNC in Sect.~\ref{parspace}.

 \clearpage

\section{Main formation and destruction pathways for different values of $\zeta$}
\label{chemicalnetworks}
\begin{figure*}
\centering
\includegraphics[width=45pc]{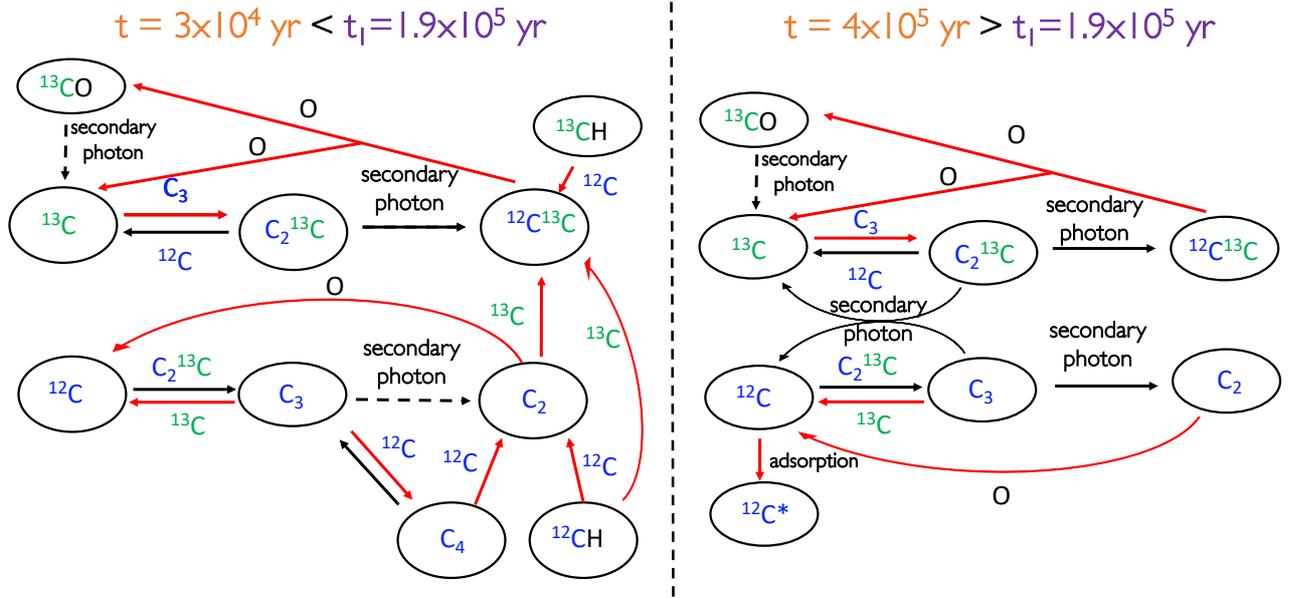}
\caption{Chemical pathways that distribute the two carbon isotopes in atomic carbon, C$_{2}$ and C$_{3}$ at 3$\times$10$^{4}$ yr (\emph{left panel}) and 4$\times$10$^{5}$~yr (\emph{right panel}), for the fiducial model with a cosmic-ray ionization rate of 1.3$\times$10$^{-18}$ s$^{-1}$. Main creation and destruction reactions are highlighted in red, $^{12}$C is represented in blue and $^{13}$C is represented in green. In this plot the dashed black reactions are the less important ones.}
\label{fig-13C-reactions-lowerz}
\end{figure*}

\begin{figure*}
\centering
\includegraphics[width=45pc]{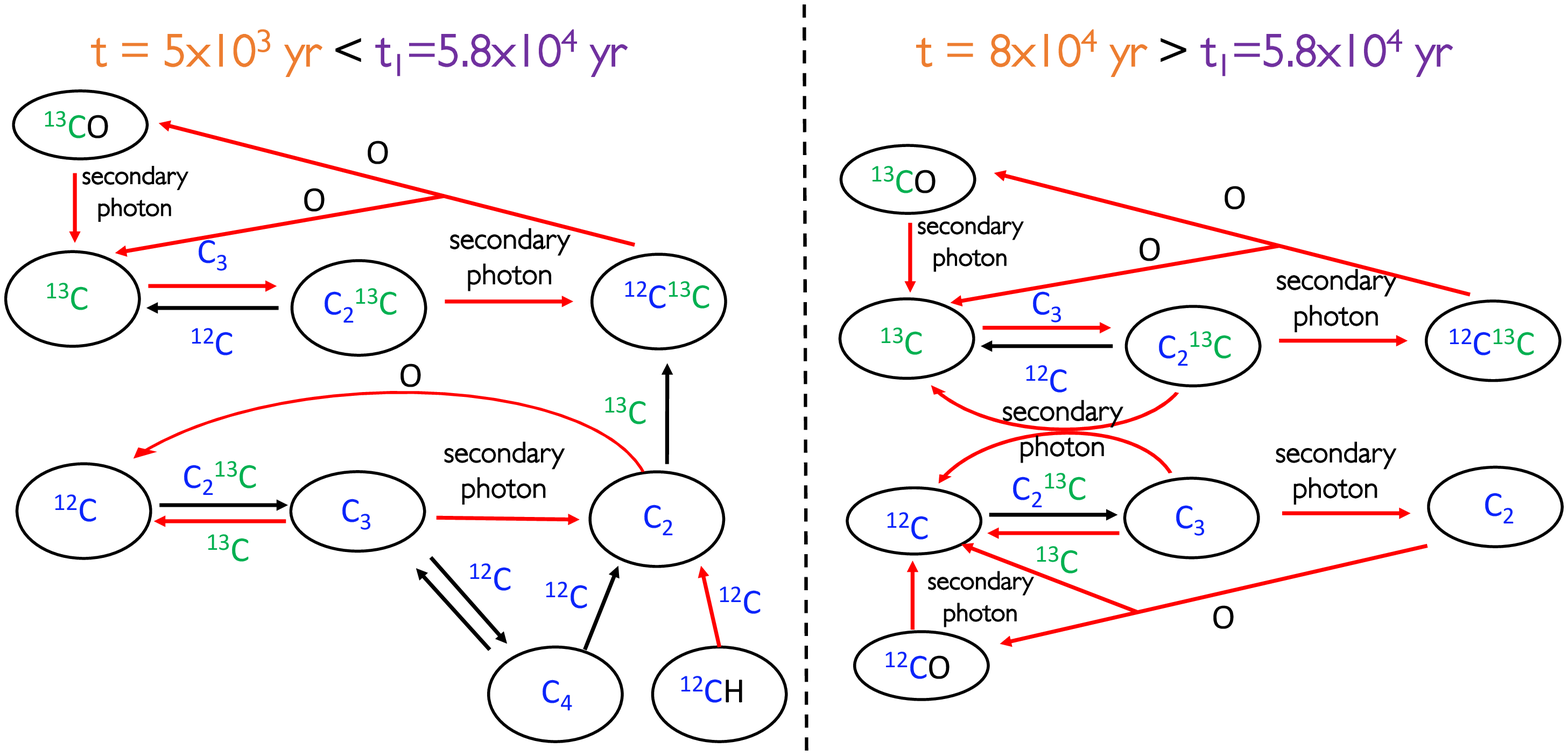}
\caption{Chemical pathways that distribute the two carbon isotopes in atomic carbon, C$_{2}$ and C$_{3}$ at 5$\times$10$^{3}$ yr (\emph{left panel}) and 8$\times$10$^{4}$~yr (\emph{right panel}), for the fiducial model with a cosmic-ray ionization rate of 1.3$\times$10$^{-16}$ s$^{-1}$. Main creation and destruction reactions are highlighted in red, $^{12}$C is represented in blue and $^{13}$C is represented in green.}
\label{fig-13C-reactions-higherz}
\end{figure*}

In this Appendix the main destruction and formation pathways of various species for $\zeta_{\rm low}$ and $\zeta_{\rm high}$ are shown (Figs.~\ref{fig-13C-reactions-lowerz} and \ref{fig-13C-reactions-higherz}, respectively). The main discussion about the different results obtained assuming a different $\zeta$ is given in Sect.~\ref{effectCR}.

 \end{appendix}
\end{document}